\documentclass[aps,prd,prabib,showpacs,nofootinbib]{revtex4}
\usepackage{graphicx} \usepackage{amsmath} \usepackage{amssymb}
\usepackage{amsfonts} \usepackage{bm}
\usepackage{array}
\usepackage{siunitx}
\usepackage[singlelinecheck=false]{caption}
\usepackage{ytableau}
\usepackage{multirow}

\newcommand*{\Comb}[2]{{}^{#1}C_{#2}}

\begin{document}

\newcommand{\be}{\begin{equation}} \newcommand{\ee}{\end{equation}}
\newcommand{\bea}{\begin{eqnarray}}\newcommand{\eea}{\end{eqnarray}}

\title{Heisenberg  Model and Rigged Configurations}

\author{Pulak Ranjan Giri} \email{pulakgiri@gmail.com}

\author{Tetsuo Deguchi} \email{deguchi@phys.ocha.ac.jp}

\affiliation{ Department of Physics, Graduate School of Humanities and Sciences, Ochanomizu University, Ohtsuka 2-1-1, Bunkyo-ku, Tokyo, 112-8610, Japan}

\begin{abstract}
We show a  correspondence of all the   solutions   of  the spin-$1/2$ isotropic  Heisenberg  model for $N=12$   to  the rigged configurations   based on the comparison  of the  set of Takahashi quantum numbers in lexicographical order  with the set of  riggings of  the  rigged configurations in co-lexicographical order.
\end{abstract}


\date{\today}

\maketitle

\section{Introduction}

The  study of  the isotropic  spin-$1/2$   Heisenberg  model     still remains  to be very much important   in the research field  of the integrable models  due to its  complex  structure  of  solutions, which  have not been completely  understood till to date.  Although  the eigenvalue equation  was solved several decades  ago   \cite{bethe},  the  completeness  issue of the spectrum  still remains  an open  problem.  For a deeper understanding  of  the
isotropic  spin-$1/2$   Heisenberg  model, its variants  and for various approaches   to solve these models    see   refs.    \cite{gaud,faddeev,korepin,faddeev1,takab,nepo2,koma,mart,bei,suth,vlad2,woy,bab,fab1,fab2}.  Getting the solutions to the Bethe ansatz equations, which produce eigenvalues and their  corresponding Bethe eigenstates,    are difficult  because  of their   high degree of nonlinearity and multi-variate   nature.  Numerical method is  the  only  way, where  several attempts  have been made to obtain the solutions  to the Bethe ansatz equations. First   complete  solutions for $N=8, 10$   spin-$1/2$ XXX chains  were  obtained in  ref. \cite{hag}, where the Bethe ansatz equations are solved by an iteration method.  On the other hand  in  ref.  \cite{nepo1} (see supplemental material)  solutions  up to $N=14$ are obtained by  making use of   the homotopy continuation  method.

The solutions to the Bethe ansatz equations have a rich and  diverse structure. As known to Bethe himself, the  complex solutions  come in complex conjugate pairs  and typically arrange themselves in a  set of string like structures.   This apparent distribution of the rapidities in the form of  strings  leads to  the   {\it string hypothesis} \cite{taka}.  The strings  together with their  deformations  are  very much  useful for the study of  the  finite length  spin chains.  Although it is known that  the     {\it string hypothesis} does not work for  some solutions of  relatively large yet finite spin chains due to the collapse of certain numbers to string solutions, it still works well in practical purpose for most of the solutions of the finite length  chains.   The complex solutions, which are responsible for the formation of bound states, present  more challenges in the numerical methods compared to the real solutions.  Nonetheless   the  knowledge of  the complex  solutions  are  essential for the understanding  of the  dynamics of  the model and for  the evaluations   of  the   correlation functions \cite{maillet,maillet1} and other physical quantities.
Recently it  has been shown \cite{giri} that  some of the solutions of  $N=12$, which have  some odd-length strings with complex centers,  do not  fit into the standard form of the  string solutions. The existence of the  complex centered odd-length strings make those strings non-self-conjugate individually,  although  the solutions as a whole  remain self-conjugate. 
Another interesting solutions are  the singular solutions, which,  because of  their  two constituent rapidities of the form  $\pm \frac{i}{2}$,   make   the eigenvalues and their corresponding Bethe eigenstates   ill-defined.  A suitable regularization scheme \cite{vlad1,sid,nepo,nepo1} is necessary  to obtain   well-defined eigenvalues and their  corresponding Bethe eigenstates for      
these type of solutions. 

The  different classes  of solutions make the   counting  process, known as the completeness  of  the spectrum,  very  difficult.  Although  {\it string hypothesis}  provides  a proof for the completeness, its  certain assumptions  are not  legitimate  \cite{vlad,essler,isler,ila,fuji}  for the  finite length chains.   One way out is to check the completeness   by obtaining all the solutions of the Bethe ansatz equations.    On the other hand,  there is a  nice  one-to-one  and onto  mapping between the solutions of the Bethe ansatz equations and  a  combinatorial  object known as the  rigged configuration  \cite{kirillov1,kirillov3,kirillov4,lule}, which is helpful   to understand the completeness problem.  However,   there are certain solutions for the large length chains, such as  the solutions corresponding to the  collapse of   the   $2$-strings, which do not have straight forward mapping to the rigged  configurations.  Recent conjecture  \cite{kirillov}   on  the rigged configurations  has    predicted   the numbers of singular solutions  present  in a  spin chain.  Although, the  correspondence between  a few  solutions of the Bethe ansatz equations  for some lengths of the spin chain to the rigged configurations  are  known, the complete  correspondences  of all the solutions of a spin chain  has  never been  shown.

The purpose of this present paper  is to provide a complete correspondence  between all  the solutions of the  Bethe ansatz equations of an isotropic spin-$1/2$ Heisenberg  chain  for $N=12$ and the rigged configurations.   The reason for  choosing   $N=12$  case  is the following:  For an even-length spin chain,    $N=12$  is the lowest  length which has some intriguing features  in the  solutions of the  Bethe ansatz  equations.   One is, as mentioned  above, there  are  some  roots, called non-self-conjugate string solutions, which do not fit in the standard framework  of  the 
{\it string hypothesis}.   The other being,  which we will show  in this paper,  there are some solutions which  seem  to  be of the  form   $ \lambda  \pm 0.5 i$.   However,  numerical evaluations with higher  working precision  show that  there are small  imaginary deviation  $\epsilon^N i$ to these solutions, which  are  usually overlooked in the numerical  methods.   Note that $N=12$  can be   directly  diagonalized   to obtain the eigenvalues. Recently  it has been  completely solved by   homotopy continuation  method  \cite{nepo1}, where the above mentioned two  types of solutions   have not be given  proper  attention.   We alternatively   study   the $N=12$  spin  chain   by an iteration method making use of the   
deformed strings  structures of the solutions  and obtain  a  correspondence   among the  solutions to the   Bethe quantum numbers, Takahashi quantum numbers and  the rigged configurations.   One advantage of this method is that  it enables us to obtain  the  Bethe quantum numbers and   also the Takahashi quantum numbers, which   are compared to the riggings of a rigged configuration to obtain the  mapping.

We organize this paper in the following fashion:  In  section II,   we review the  the algebraic Bethe ansatz method for the  spin-$1/2$ isotropic Heisenberg  model.  In section III we  discuss the string solutions which include  non self-conjugate string solutions,  {\it almost singular}  string solutions and  singular string solutions. 
In section IV we  study   the   correspondence of the rigged configurations    to  the  $N=12$   spin-$1/2$  isotropic Heisenberg model  based on a comparison of  the Takahashi quantum numbers with the riggings   and finally we conclude.


\section{Algebraic Bethe Ansatz Method}
In the algebraic Bethe ansatz method  the Lax operator for the   spin-$1/2$  isotropic Heisenberg  model  is given by 
\begin{eqnarray}
L_\gamma(\lambda)=\left( \begin{array}{cc}
\lambda-iS^z_\gamma & -iS^-_\gamma  \\
-iS^+_\gamma & \lambda+iS^z_\gamma  \end{array} \right)\,,
\end{eqnarray}
where $S^\pm_\gamma= S^x_\gamma \pm iS^y_\gamma$,  $S_i^j (j=x,y,z)$  is   the  spin-$1/2$ operator at the  $i$-th lattice site  and in $j$-direction.
Each element  of $L_\gamma(\lambda)$ is  a matrix of dimension  $2^N \times 2^N$, which  acts nontrivially  on the $\gamma$-th lattice site.  The monodromy matrix, $T(\lambda)$, is then given by the  direct product of  the Lax matrices  at each site 
\begin{eqnarray}\label{monodromy}
T(\lambda)=L_N(\lambda)L_{N-1}(\lambda) \cdots L_1(\lambda)=\left( \begin{array}{cc}
A(\lambda)& B(\lambda)  \\
C(\lambda) & D(\lambda) \end{array} \right)\,.
\end{eqnarray}
The transfer matrix is  obtained from the monodromy matrix    (\ref{monodromy}) as 
\begin{eqnarray}\label{tmatrix}
t(\lambda)=  A(\lambda)+D(\lambda)\,.
\end{eqnarray} 
The Hamiltonian of  the  spin-$1/2$  isotropic Heisenberg model  on a one-dimensional periodic lattice  of  length $N$  is then obtained  by taking the   logarithm  of   $t(\lambda)$ at  $\lambda= \frac{i}{2}$ as 
\begin{eqnarray}
H=  \frac{J}{2}\left(i\left[\frac{d}{d\lambda}\log t(\lambda)\right]_{\lambda=\frac{i}{2}}-N\right) = J\sum_{i=1}^{N}\left(S^x_iS^x_{i+1}+S^y_iS^y_{i+1}+S^z_iS^z_{i+1}- \frac{1}{4}\right)\,.
\end{eqnarray} 
The Bethe states for    $M$ down-spins  are given by 
\begin{eqnarray}\label{vec0}
|\lambda_1, \lambda_2,\cdots,\lambda_M\rangle= \prod_{\alpha=1}^MB(\lambda_\alpha)|\Omega\rangle\,,
\end{eqnarray}
where  $|\Omega\rangle$  is the reference eigenstate with all  up-spins,   $B(\lambda_\alpha)$   is  an element of  the monodromy matrix $T(\lambda_\alpha)$ obtained from  eq. (\ref{monodromy}) and  $\lambda_\alpha$ are the rapidities. 
More explicitly the Bethe  state (\ref{vec0}) can   be  expressed   as  \cite{deguchi1}
\begin{eqnarray}\label{vec1}\nonumber
\prod_{\alpha=1}^MB(\lambda_\alpha)|\Omega\rangle= &&(-2i)^M \prod_{j < k}^{M}\frac{\lambda_j-\lambda_k +i}{\lambda_j-\lambda_k} \prod_{j =1}^{M}\frac{(\lambda_j- \frac{i}{2})^N}{\lambda_j+ \frac{i}{2}}\times \\
&&\sum_{1\leq x_1 < x_2 \cdots < x_M \leq N}^{N}\sum_{\mathcal{P}\in S_M}^{M!} \prod_{\mathcal{P}j <\mathcal{P}k}^{M} \left(\frac{\lambda_{\mathcal{P}j}-\lambda_{\mathcal{P}k}-i}{\lambda_{\mathcal{P}j}-\lambda_{\mathcal{P}k}+i}\right)^{H(j-k)} \prod_{j=1}^M\left(\frac{\lambda_{\mathcal{P}j}+ \frac{i}{2}}{\lambda_{\mathcal{P}j}- \frac{i}{2}}\right)^{x_j}\prod_{j=1}^M S^-_{x_j}|\Omega \rangle\,,
\end{eqnarray}
where $S_M $  is the permutation group  of $M$ numbers  with elements  $\mathcal{P}$  and the  Heaviside step function  is defined as  $H(x) =1$ for $x >0$ and  $H(x) =0$ for $x \leq 0$.   
Acting the transfer matrix   $t(\lambda)$ from the left on the Bethe state  (\ref{vec0})   we obtain 
\begin{eqnarray}\label{tb}
t(\lambda)\prod_{\alpha=1}^MB(\lambda_\alpha)|\Omega\rangle= \Lambda\left(\lambda, \{\lambda_\alpha\}\right)\prod_{\alpha=1}^MB(\lambda_\alpha)|\Omega\rangle + \sum_{k=1}^{M} 
\Lambda_{k}\left(\lambda, \{\lambda_\alpha\}\right)B(\lambda)\prod_{\alpha \neq k}^MB(\lambda_\alpha)|\Omega\rangle\,,
\end{eqnarray}
where   eigenvalue of the transfer matrix is given by 
\begin{eqnarray}\label{teigen}
\Lambda\left(\lambda, \{\lambda_\alpha\}\right)= \left(\lambda + \frac{i}{2}\right)^N \prod_{\alpha=1}^M\frac{\lambda-\lambda_\alpha-i}{\lambda-\lambda_\alpha} + 
\left(\lambda - \frac{i}{2}\right)^N \prod_{\alpha=1}^M\frac{\lambda-\lambda_\alpha+i}{\lambda-\lambda_\alpha}\,,
\end{eqnarray}
and the  unwanted terms are given by 
\begin{eqnarray}\label{tunw}
\Lambda_k\left(\lambda, \{\lambda_\alpha\}\right)= \frac{i}{\lambda-\lambda_k}\left[\left(\lambda_k + \frac{i}{2}\right)^N \prod_{\substack{{\alpha =1} \\{\alpha\neq k}}}^M\frac{\lambda_k-\lambda_\alpha-i}{\lambda_k-\lambda_\alpha} + 
\left(\lambda_k - \frac{i}{2}\right)^N \prod_{\substack{{\alpha =1} \\{\alpha\neq k}}}^M\frac{\lambda_k-\lambda_\alpha+i}{\lambda_k-\lambda_\alpha}\right]\,,~ k= 1, 2, \cdots,  M\,.
\end{eqnarray}
Eq. (\ref{tb}) reduces to  the  eigenvalue equation, if  the unwanted terms  (\ref{tunw}) vanish, which  enables us to obtain the   Bethe ansatz equations 
\begin{eqnarray}\label{bethe0}
\left(\frac{\lambda_\alpha- \frac{i}{2}}{\lambda_\alpha+ \frac{i}{2}}\right)^{N}= \prod_{\substack{{\beta =1} \\{\beta\neq \alpha}}}^{M}\frac{\lambda_\alpha-\lambda_\beta-i}{\lambda_\alpha-\lambda_\beta+i}\,, ~~~  \alpha=1,2, \cdots, M\,.
\end{eqnarray}
Once the solutions $\lambda_\alpha$ of the   Bethe ansatz equations   (\ref{bethe0}), known as the Bethe roots,   are   obtained,  the eigenvalues  of the Hamiltonian  $H$ for  the  $M$ down-spin sector  are  expressed  as   
\begin{eqnarray}\label{eigen}
E=   \frac{J}{2}\left(i\left[\frac{d}{d\lambda}\log \Lambda\left(\lambda, \{\lambda_\alpha\}\right)\right]_{\lambda=\frac{i}{2}}-N\right)= -J\frac{1}{2}\sum_{\alpha=1}^{M}\frac{1}{\left(\lambda_\alpha^2+ \frac{1}{4}\right)}\,.
\end{eqnarray}
The logarithmic form of  eq. (\ref{bethe0})   is  expressed as 
\begin{eqnarray}\label{logform}
2\arctan(2\lambda_\alpha)= J_\alpha\frac{2\pi}{N} + \frac{2}{N}\sum_{\substack{{\beta =1}\\{\beta\neq\alpha}}}^{M}\arctan(\lambda_\alpha-\lambda_\beta)\,, ~~~  \alpha=1,2, \cdots, M\,, ~~~~\mbox{mod}~2\pi\,,\end{eqnarray}
where   $\{J_\alpha, \alpha=1,2, \cdots, M\}$ are the Bethe quantum numbers, which   take  integral (half integral)  values   if   $N-M$ is odd (even)  respectively.    Since the Bethe quantum numbers are  repetitive,  they   are  not  useful for the process of  counting  the total number of states.
Therefore a strictly non-repetitive  quantum numbers are necessary, which  are obtained as follows. 

In  {\it string hypothesis}, the rapidities of a solution  to  the Bethe ansatz equations are arranged in form of strings of different lengths as 
\begin{eqnarray}\label{string}
\lambda_{\alpha a}^{j}= \lambda_{\alpha}^{j} +  \frac{i}{2}\left(j+1-2a\right) +  \Delta_{\alpha a}^{j}\,, ~~~ a=1,2, \cdots, j, ~~\alpha=1,2,.., M_j\,,
\end{eqnarray}
where the   $j$-string of length $j$ has the real center   $\lambda_\alpha^j$ with   $\alpha$  being the index to distinguish  all the $M_j$  strings of same length and 
$\Delta_{\alpha a}^{j}$  are  the string deviations.  
In the limit that these  deviations vanish,  $\Delta_{\alpha a}^{j} \to 0$, equations (\ref{bethe0}) reduce to the   convenient form
\begin{eqnarray}\label{betheta}\nonumber
\arctan\frac{2\lambda^j_\alpha}{j} &=& \pi \frac{I^j_\alpha}{N} + \frac{1}{N}\sum_{k=1}^{N_s}\sum_{\beta}^{M_k}\Theta_{jk}\left(\lambda_\alpha^j-\lambda_\beta^k\right)\,,~~~~ \mbox{mod} ~\pi\,,\\
\Theta_{jk}(\lambda) &=& (1-\delta_{jk})\arctan\frac{2\lambda}{|j-k|} + 2\arctan\frac{2\lambda}{|j-k|+2} + \cdots + 2\arctan\frac{2\lambda}{j+k-2} + \arctan\frac{2\lambda}{j+k}\,,
\end{eqnarray}
where  $M$ down-spins are partitioned into  $M_k$    $k$-stings with the length of the largest string being  $N_s$  such that  $\sum_{k}^{N_s}kM_k=M$.   We now obtain the strictly non-repetitive quantum numbers   $I^j_\alpha$, known as the Takahashi quantum numbers,  from eq.   (\ref{betheta}), which     have the following bounds
\begin{eqnarray}\label{takahashi} 
\mid I_{\alpha}^j\mid \leq \frac{1}{2} \left(N-1-\sum_{k=1}\left[2 \mbox{min}(j,k)-\delta_{j,k}\right]M_k\right)\,.
\end{eqnarray}

\section{String solutions}
The solutions of the Bethe ansatz equations  are usually arranged in a set of strings.   In  {\it string hypothesis} language the real solutions are   $1$-string solutions.  Other solutions are composed of strings of different lengths as given in eq.  (\ref{string}).  In the numerical iteration method  first the   centers of the strings      $ \lambda_{\alpha}^{j}$ are  evaluated, which are used as the initial input for the next iteration method to  calculate the  deviations  $\Delta_{\alpha a}^{j}$  and the centers.   For finite  but not so large spin chains,   the deviations of the strings  are of the  form  \begin{eqnarray}\label{stringdev}
\Delta_{\alpha a}^{j}=  \epsilon_{\alpha a}^j + i\delta_{\alpha a}^j\,,
\end{eqnarray}
where  the real $\epsilon_{\alpha a}^j $ and  $\delta_{\alpha a}^j$  decreases  exponentially  with  respect the length  $N$ of the spin chain. 
While implementing  the iteration scheme,  the  complex conjugacy condition of the solutions are  exploited to   simplify  the  forms  of these  deviations.  One useful way to implement it is to consider the strings  to be individually self-conjugate, which implies \cite{hag} 
\begin{eqnarray}\label{stringdev1}
\Delta_{\alpha a}^{j}= (\Delta_{\alpha  j+1-a}^{j})^*\,.
\end{eqnarray}
The centre for the odd-length  strings  with these  conditions   are  inevitably  real.  Since each  of the stings in a given solution  is  self-conjugate, they are not related to each other.    For the even-length spin chains  up to $N=10$  all the solutions  obey   the conditions  (\ref{stringdev1}). Even most of the solutions for  $N=12$  given in  {\bf Tables}  $1-6$ also obey these conditions.  However  a small number of solutions for $N=12$  do not  obey  the self-conjugacy  conditions  (\ref{stringdev1}), which we discuss in the next subsection. 


\subsection{Non Self-conjugate  String solutions}
There are a few number of solutions for $N=12$ spin-$1/2$ chain, which do not fall in the standard form of the sting solutions discussed above.  Note  that in a scheme of small deviations, the self-conjugacy conditions  (\ref{stringdev1}) are too restrictive to be valid for all solutions.   It is therefor necessary to impose the self conjugacy condition on  whole set of the   rapidities in a solution
\begin{eqnarray}\label{conju1}
 \{\lambda_{\alpha a}^{j}\} = \{(\lambda_{\alpha a}^{j})^*\} \,.
\end{eqnarray}
Note that  (\ref{conju1})  implies   (\ref{stringdev1}), however the reverse  is not true always. 
Relaxing the self-conjugacy condition as   (\ref{conju1}) allows  the centre of the  odd-length strings  to  be complex with small imaginary  part in such a way that  the  solutions  remain  self-conjugate. 
These  type of complex centered odd-string solutions, we call {\it non self-conjugate string}  solutions,   have been observed to appear for even-length spin chains   for   $ N \geq 12$.  Here we remark  that  it is essential to  keep the deviations small, so that the    $\Delta_{\alpha a}^{j} \to 0$ limit  lead  to the  Takahashi  equations  (\ref{betheta}) and  the Takahashi quantum numbers  (\ref{takahashi}) can be obtained. 

The first two examples  of such   {\it non self-conjugate string}  solutions appear  in a set of solutions with  two 1-strings and one 3-string  in  $M=5$ down-spins sector   of the   $N=12$ spin chain.
The self-conjugacy conditions  (\ref{stringdev1})  allow  us to write the   rapidities of the solutions   in a simple form as   
 \begin{eqnarray}\label{5downspin}
 \{\lambda_1\},\{\lambda_2\},\{\lambda+\epsilon + (1+\delta)i,\lambda, \lambda+\epsilon - (1+\delta)i\}\,,
\end{eqnarray}
where the  real parameters $\lambda_1,\lambda_2,\lambda$  are the centers and $\epsilon,\delta$ are small  deviations.   The curly brackets in the above expressions   mean a  set of rapidities  of a given sting.  In  {\bf Table $5.3$}  all solutions expect two   are obtained  making use of    (\ref{5downspin}) in the iteration   process.   For the two   {\it non self-conjugate string}  solutions we make use of the  relaxed self-conjugacy condition   (\ref{conju1}), which allows  us to write  the rapidities as 
\begin{eqnarray}\label{5downspin1}
 \{\lambda_1\},\{\lambda +i\delta_1\},\{\lambda+\epsilon + (1+\delta_2)i,\lambda-i\delta_1, \lambda+\epsilon - (1+\delta_2)i\}\,,
\end{eqnarray}
where  the real parameters $\lambda_1,\lambda$ are  the centers and $\epsilon,\delta_1,\delta_2$ are  the deviations.   The  solutions   $171$ and  $172$  in  {\bf Table  $5.3$}  are obtained   by making use of (\ref{5downspin1}) in the iteration process.  Notice that because of the imaginary term  $i\delta_1$ the  $1$-string   $\{\lambda +i\delta_1\}$  and the  $3$-string  
$\{\lambda+\epsilon + (1+\delta_2)i,\lambda-i\delta_1, \lambda+\epsilon - (1+\delta_2)i\}$  are individually  non self-conjugate although together   remain  self-conjugate.

The next three  examples    of such   {\it non self-conjugate string}  solutions appear in a set of solutions with  one 1-string, one 2-string  and one 3-string in $M=6$ down-spins sector of the  $N=12$ spin chain.  The self-conjugacy conditions  (\ref{stringdev1})  allow  us to write the   rapidities  of the solutions  in a simple form as   
\begin{eqnarray}\label{6downspin}
 \{\lambda_2\},\{\lambda_1 +\frac{i}{2}(1+2\delta), \lambda_1 -\frac{i}{2}(1+2\delta)\},\{\lambda+\epsilon + (1+\delta_1)i,\lambda, \lambda+\epsilon - (1+\delta_1)i\}\,,
\end{eqnarray}
where the real  parameters $\lambda_1,\lambda_2,\lambda$  are the centers and  the real parameters $\delta,\delta_1,\epsilon$ are small deviations.  In  {\bf Table $6.8$}  all solutions expect three   are obtained  making use of    (\ref{6downspin}) in the iteration   process.   For the three   {\it non self-conjugate string}  solutions we make use of the  relaxed self-conjugacy condition   (\ref{conju1}), which allows  us to write  the rapidities as 
\begin{eqnarray}\label{6downspin1}
 \{\lambda +i\delta_1\},\{\lambda_1 +\frac{i}{2}(1+2\delta), \lambda_1 -\frac{i}{2}(1+2\delta)\},\{\lambda+\epsilon + (1+\delta_2)i,\lambda-i\delta_1, \lambda+\epsilon - (1+\delta_2)i\}\,,
\end{eqnarray}
where the real parameters  $\lambda,\lambda_1, \delta_2$  are the centers and  the real parameters  $\delta, \delta_1, \epsilon$ are  small deviations.    The  solutions   $105, 118$ and  $119$  in  {\bf Table  $6.8$}  are obtained   by making use of (\ref{6downspin1}) in the iteration process.  The presence of the imaginary term  $i\delta_1$  makes the $1$-string  $ \{\lambda +i\delta_1\}$  and the 
$3$-string   $\{\lambda+\epsilon + (1+\delta_2)i,\lambda-i\delta_1, \lambda+\epsilon - (1+\delta_2)i\}$  non self-conjugate individually although  together they remain self-conjugate. 


\subsection{Almost singular string solutions}

There are some solutions  $\{\lambda_1, \lambda_2, \lambda_3, \cdots \lambda_M\}$ for $N=12$ spin-$1/2$ chain,  in which a pair of rapidities  are of the form 
\begin{eqnarray}\label{asingular}
{\lambda}_1 = a\epsilon + \frac{i}{2}\left(1 + 2\epsilon^N\right)\,,~~~
{\lambda}_2 = a\epsilon - \frac{i}{2}\left(1 + 2\epsilon^N\right)\,,
\end{eqnarray}
where    $a$  and $\epsilon$  are small numbers such that  $a\epsilon$ is real in the case of $N=12$.    $\epsilon^N $ is so small that  in  numerical calculation  the contribution  of   $\epsilon^N i$  does not usually appear unless the working precision  is increased   significantly.   Most calculations  wrongly  produce  solutions,  in which   the above  pair takes the form 
\begin{eqnarray}\label{asingular1}
{\lambda}_1 = a\epsilon + \frac{i}{2}\,,~~~
{\lambda}_2 = a\epsilon - \frac{i}{2}\,, 
\end{eqnarray}
as  obtained  in  the supplements of ref.  \cite{nepo1}.   However,  analytically  also   it can be argued that  there does  not exist   any solution  containing  a pair of rapidities of the form  (\ref{asingular1}).      Because  of the similarity   with the   two regularized  rapidities of the singular solutions   with  (\ref{asingular}),  we call  such  solutions   {\it  almost singular}.    
For  $M=4$ down-spins  sector  there are two such solutions   $257, 258$  with two  $2$-strings  shown  in {\bf Table $4.4$}.   
For $M=5$ down-spins sector there are six such   solutions,  two solutions  $48, 49$   with three $1$-strings and one $2$-string, which are shown  in  { \bf Table $5.2$},   two  solutions  $212, 213$   with  one $1$-string and one $4$-string shown in {\bf Table $5.4$}   and   the  last two  solutions   $242, 243$   with  one $1$-string and two $2$-strings shown  in { \bf Table $5.6$}.    
For $M=6$ there are  two such solutions $55, 56$  with  two  $1$-strings and one  $4$-string    shown in  { \bf Table $6.4$}.  Note that {\it almost singular}  solutions obtained for $N=12$ are all self-conjugate string solutions. 


\subsection{Singular string solutions}
There are some solutions, known as the  singular string  solutions,  of  the form 
\begin{eqnarray}\label{sgnrap}
\Big\{\lambda_1=\frac{i}{2},\lambda_2= -\frac{i}{2}, \lambda_3,\lambda_4, \cdots, \lambda_M \Big\}\,,
\end{eqnarray} 
which produce  ill-defined  eigenvalues and the Bethe eigenstates  due to the presence of the two rapidities  of the form  $\lambda_1=\frac{i}{2},\lambda_2= -\frac{i}{2}$.  To obtain  finite and well-defined  eigenvalues and the Bethe eigenstates  for these solutions it is imperative  to  exploit a suitable regularization scheme   \cite{bei,vlad1,nepo,sid,kirillov2,girir}.   For the even length  spin chains it has been observed that the rapidities and  the corresponding Takahashi quantum numbers  are distributed symmetrically   making the sum of the rapidities to  vanish 
\begin{eqnarray}\label{singcond}
\sum_{\alpha=1}^M\lambda_\alpha= 0\,.
\end{eqnarray}
Note that the condition  (\ref{singcond})   is not only satisfied by singular string solutions, but also by   other solutions, which  also are invariant  under the negation of their  rapidities.   The condition 
(\ref{singcond}), noted in \cite{hag},  forms a part  of a conjecture in  ref.   \cite{kirillov}.

The simplest singular  string solution, found   in $M=2$ down-spins sector,  is of the form  
\begin{eqnarray}\label{2sol1}
\Big\{\frac{i}{2},-\frac{i}{2}\Big\}\,.
\end{eqnarray}
From  condition   (\ref{singcond}) it  clear  that  there is one singular solution in this sector, which is given    as solution $46$  with one $2$-string in   {\bf Table  $2.2$}.    In $M=3$ down-spins  sector also condition    (\ref{singcond})   gives   one solution of the form 
\begin{eqnarray}\label{3sol1}
\Big\{\frac{i}{2},-\frac{i}{2}, 0\Big\}\,.
\end{eqnarray}
In {\bf Table $3.2$}  such solution is given in  $85$ with one $1$-string and one $2$-string.   In $M=4$ down-spins sector two class of solutions, compatible with the condition   
(\ref{singcond}),  are given by
\begin{eqnarray}\label{4sol1}
&&\Big\{\frac{i}{2},-\frac{i}{2}, a_1, -a_1\Big\} ~~~~~~ \mbox{for}~~~ a_1 \in \mathbb{R}\,,\\ \label{4sol2}
&&\Big\{\frac{i}{2},-\frac{i}{2},  ia_1, -ia_1\Big\} ~~~~ \mbox{for}~~~ a_1\in \mathbb{R}\,. 
\end{eqnarray}
The solutions  $71, 76,  91,  98$   in {\bf Table  $4.2$}   with  two $1$-strings and one $2$-string  are of the form  (\ref{4sol1}) and the solution  $271$  in {\bf Table $4.5$}
with one $4$-string  is of the form    (\ref{4sol2}). 
In $M=5$ down-spins   sector there are two classes of singular sting solutions of the form
\begin{eqnarray}\label{5sol1}
&&\Big\{\frac{i}{2},-\frac{i}{2}, 0,a_1, -a_1\Big\}~~~~~~ \mbox{for}~~~ a_1 \in \mathbb{R}\,,\\ \label{5sol2}
&&\Big\{\frac{i}{2},-\frac{i}{2}, 0, ia_1, -ia_1\Big\}~~~~ \mbox{for}~~~ a_1\in \mathbb{R}\,.
\end{eqnarray}
The solutions  $22, 29,  50$  in {\bf Table $5.2$} with three $1$-strings and  one $2$-string  are of the form  of  (\ref{5sol1}).  Whereas  the solution  $211$ in {\bf Table $5.4$} with one $1$-string and one $4$-string  and the solution   $283$  in {\bf Table $5.7$}  with one $2$-string and one $3$-string are of the form  of  (\ref{5sol2}). 
In $M=6$ down-spins sector   there are four classes of solutions of the form 
\begin{eqnarray}\label{sol1}
&&\Big\{\frac{i}{2},-\frac{i}{2}, a_1, -a_1, a_2, -a_2\Big\}~~~~~~~ \mbox{for}~~~ a_1, a_2 \in \mathbb{R}\,,
\\ \label{sol2}
&&\Big\{\frac{i}{2},-\frac{i}{2}, a_1, -a_1, ia_2, -ia_2\Big\}~~~~~ \mbox{for}~~~ a_1, a_2 \in \mathbb{R}\,,
\\ \label{sol3}
&&\Big\{\frac{i}{2},-\frac{i}{2}, ia_1, -ia_1, ia_2, -ia_2\Big\} ~~~ \mbox{for}~~~ a_1, a_2 \in \mathbb{R}\,,
\\ \label{sol4}
&&\Big\{\frac{i}{2},-\frac{i}{2}, a_1\pm ia_2, -a_1 \pm ia_2\Big\}~~~ \mbox{for}~~~ a_1, a_2 \in \mathbb{R}\,.
\end{eqnarray}
The solutions  $2, 11, 16$   in {\bf Table $6.2$}  with  four $1$-strings and one $2$-string are of the form   (\ref{sol1}).   The solutions  $52, 57, 72, 79$   in {\bf Table $6.4$}  with  two $1$-strings and 
one $4$-string are of the form     (\ref{sol2}).     The solution  $89$ in  {\bf Table $6.6$} with one $6$-string   and  the solution  $105$ in  {\bf Table $6.8$}  with one $1$-string,  one  $2$-string and one $3$-string   are of the form   (\ref{sol3}).   And finally the solution  $126$  in   {\bf Table $6.9$}  with three $2$-strings are of the form  (\ref{sol4}). 


\section{Rigged Configurations}
For finite length  spin chains with no collapsing strings  there is a  correspondence between  the  solutions of the Bethe ansatz equations  and a combinatorial object, known as   the rigged configurations \cite{kirillov,kirillov2}.
Although,   this correspondence  has been shown for a few solutions of  some given length spin-$1/2$ chains, the complete  bijection of all the solutions of the Bethe ansatz equations to the rigged configurations have never been shown.   Since  this  knowledge of correspondence  is  very  much  important  in  the completeness  of  the  solutions, we in this section discuss the  the bijection  based on the comparison of the Takahashi quantum numbers  with   the riggings of the rigged configurations.

Before we  study  a correspondence between the solutions of the Bethe ansatz equations to the rigged configurations  let us  briefly formulate    the rigged configurations  first.  These are the  Young Tableau like objects, which have two sets of integers  to  label them.  The numbers on the left hand sides of the boxes are known as  the  vacancy numbers $P_k(\nu)$  and the numbers on the right hand sides of the boxes are known as the  riggings $J_{k,\alpha}$.  In order to have these numbers,  first the total numbers of down-spins  $M$ in a given  solution   is  divided into  $s$ parts as 
$\nu= \{\nu_1, \nu_2, \cdots, \nu_s\}$,  such that  the parts   $\nu_i $'s  are positive and satisfy    $\sum_{i=1}^{s}\nu_i= M$.  For a spin-$1/2$ chain the   vacancy number  $P_k(\nu)$ for a string of length $k$  in   a specific partition   $\nu$  is    given by  
\begin{eqnarray}\label{rigg1}
P_k(\nu)= N- 2 \sum_{i=1}^{s} \mbox{min} \left(k,\nu_i\right)\,,
\end{eqnarray}
where  all the vacancy numbers are  non-negative.    Given a  vacancy number   the  corresponding    riggings   $J_{k,\alpha}$  are  defined  as 
\begin{eqnarray}\label{rigg2}
0\leq J_{k, 1} \leq J_{k, 2} \leq \cdots \leq J_{k, M_k}\leq P_k(\nu)\,,
\end{eqnarray}
The  flip map $\kappa$  among the riggings  in a rigged configuration is defined  as
\begin{eqnarray}\label{rigg3}
\kappa(J_{k, \alpha})= P_k(\nu)- J_{k, M_k-\alpha+1}\,,
\end{eqnarray}
which plays  the same role as the  transformation of  all the rapidities in a solution under negation.   For a correspondence  between the  solutions of the Bethe ansaz equations  and  the rigged configurations we  need   a rule  such that  a rigged configuration  can be  assigned to a particular solution.     Such assignment can be done by comparing 
the  riggings $J$ with the real parts of the  rapidities of  the solutions of the Bethe ansatz equations.   The   higher value of the real part of the  solutions to the Bethe equations  is  assigned to the  higher value of the riggings \cite{kirillov}.    For such a scheme to work one needs  to actually solve the Bethe ansatz equations to obtain  the rapidities. Moreover  sometimes the rapidities of  the solutions of the Bethe ansatz equations  are very close  to each other,  which can make  the scheme  ineffective.   We therefore  consider  another approach in which  the  comparison 
between  the riggings with the Takahashi quantum numbers are done.


\subsection{Rigged configurations and Bethe roots  correspondence}
The mapping $f$  starts from the largest   strings  to the lowest strings in decreasing order. For example, for $M=6$ down-spin sectors with  one $3$-string, one $2$-string and one $1$-string,   $\nu= \{3, 2, 1\}$,  first the rigging  $\{J_{3, 1}\}$ of   $3$-string   is compared with the Takahashi quantum number  $\{I^1_3\}$,   then the rigging  $\{J_{2, 1}\}$ of   $2$-string   is compared with the Takahashi quantum number  $\{I^1_2\}$ and finally   the rigging  $\{J_{1, 1}\}$ of   $1$-string   is compared with the Takahashi quantum number  $\{I^1_1\}$.  In this example  a specific length of string appear once. There are solutions  where  a specific length  of string  may appear more than once. Let us consider such an example  in $M=6$ down-spin sector with one $3$-string and  three $1$-strings,
$\nu= \{3, 1, 1, 1\}$.  In this case first   the rigging  $\{J_{3, 1}\}$ of   $3$-string   is compared with the Takahashi quantum number  $\{I^1_3\}$, then  the 
the riggings  $\{J_{1, 1}, J_{1,2}, J_{1,3}\}$ of  the  three $1$-strings   are  compared with the Takahashi quantum numbers  $\{I^1_1,  I^2_1, I^3_1\}$.  In a  general perspective  we have the following map
\begin{eqnarray}\label{map1}
f : \{ J_{k, 1}, J_{k, 2}, \cdots, J_{k, M_k}\}   \to   \{ I^1_{k}, I^2_{k}, \cdots, I^{M_k}_{k}\}\,.
\end{eqnarray}
When there is one string of a particular length  in a partition,  the set of    rigging  $\{J_{k,1}\}$  is arranged in  increasing order
\begin{eqnarray}\label{map2}
 \{J_{k,1}\} =  \{ J_{k,1}^1, J_{k,1}^2,  \cdots, J_{k,1}^{2 |I_k| +1}\} \,,
\end{eqnarray}
where  $J_{k,1}^i < J_{k,1}^{i+1}$. Similarly the  set of  Takahashi quantum numbers   $\{ I^1_k\}$ is also arranged in increasing  order
\begin{eqnarray}\label{map3}
 \{I_k^1\} =  \{ I_k^{1,1}, I_k^{1,2},  \cdots, I_k^{1, 2 |I_k| +1}\} \,,
\end{eqnarray}
where  $I_k^{1,i} < I_k^{1,i+1}$.  We then easily obtain a mapping  from  $\{J_{k,1}\}$  to   $\{ I^1_k\}$   as 
\begin{eqnarray}\label{map4}
 f : J_{k,1}^i   \to  I_k^{1,i} ~~~~  i= 1, 2, \cdots,  2 |I_k| +1\,.
\end{eqnarray}
When  there are more than one $k$-strings in a  partition,  $M_k \neq 1$,  then arrange  the set of riggings   in  co-lexicographical   order  \cite{sriram} as 
\begin{eqnarray}\label{map5}
\Big\{ \left(J_{k, 1}^1, J_{k, 2}^1, \cdots, J_{k, M_k}^1\right), \left(J_{k, 1}^2, J_{k, 2}^2, \cdots, J_{k, M_k}^2\right), \cdots,  \left(J_{k, 1}^{\Comb{2 |I_k|+1}{M_k}}, J_{k, 2}^{\Comb{2 |I_k|+1}{M_k}}, \cdots, J_{k, M_k}^{\Comb{2 |I_k|+1}{M_k}}\right)  \Big\}\,.
\end{eqnarray}
where any   element    $\left(J_{k, 1}^i, J_{k, 2}^i, \cdots, J_{k, M_k}^i\right),  i=1, 2,  \cdots, {\Comb{2 |I_k|+1}{M_k}}$   of the set  (\ref{map5})  is  arranged   in  weakly  increasing order defined  in   (\ref{rigg2}). 
In co-lexicographical order the elements  $i, j (i <j)$ of  (\ref{map5}) satisfy  the following relation
\begin{eqnarray}\label{map6}
\left(J_{k, 1}^i, J_{k, 2}^i, \cdots, J_{k, M_k}^i\right)  <  \left(J_{k, 1}^j, J_{k, 2}^j, \cdots, J_{k, M_k}^j\right)\,,
\end{eqnarray}
if and only if   the  $m$ elements from the right satisfy   $J_{k, l}^i= J_{k, l}^j, ~~ l= M_k, M_k-1, \cdots,  M_k- m+1$  and  $J_{k, M_k-m}^i < J_{k, M_k-m}^j$.  The  set of Takahashi quantum numbers are arranged in lexicographical order  \cite{sriram}
\begin{eqnarray}\label{map7}
\Big\{ \left(I_{k}^{1,1}, I_{k}^{2,1}, \cdots, I_{k}^{M_k,1}\right), \left(I_{k}^{1,2}, I_{k}^{2,2}, \cdots, I_{k}^{M_k,2}\right), \cdots,  \left(I_{k}^{1,{\Comb{2 |I_k|+1}{M_k}}}, I_{k}^{2,{\Comb{2 |I_k|+1}{M_k}}}, \cdots, I_{k}^{M_k,{\Comb{2 |I_k|+1}{M_k}}}\right)  \Big\}\,.
\end{eqnarray}
where any   element  is of the form    $\left(I_{k}^{1,i} <J_{k}^{2,i} < \cdots < J_{k}^{M_k,i}\right),  i=1, 2,  \cdots, {\Comb{2 |I_k|+1}{M_k}}$.
In lexicographical order the  $i, j (i <j)$  elements of  (\ref{map7}) satisfy  the following relation
\begin{eqnarray}\label{map8}
\left(I_{k}^{1,i}, I_{k}^{2,i}, \cdots, I_{k}^{M_k,i}\right)  <  \left(I_{k}^{1,j}, I_{k}^{2,j}, \cdots, I_{k}^{M_k,j}\right)\,,
\end{eqnarray}
if and only if   the  $m$ elements from the left  satisfy   $I_{k}^{l,i}= J_{k}^{l,j}, ~~ l= 0, 1, 2, \cdots, m-1$  and  $I_{k}^{m+1,i} < I_{k}^{m+1,j}$.  Now the mapping between the  elements  of  (\ref{map5})  and  (\ref{map7})  are  given by  
\begin{eqnarray}\label{map9}
f : \left(J_{k, 1}^i, J_{k, 2}^i, \cdots, J_{k, M_k}^i\right)  \to  \left(I_{k}^{1,i}, I_{k}^{2,i}, \cdots, I_{k}^{M_k,i}\right)\,,~~  i=1, 2,  \cdots, {\Comb{2 |I_k|+1}{M_k}},.
\end{eqnarray}
On the right most column of   the {\bf Tables} $1-6$  we have    shown the  rigged configurations for each of the solutions. 


\section{Conclusions}

There is a natural correspondence  between the solutions of the  Bethe ansatz equations for the  spin-$1/2$ isotropic Heisenberg  model  to the rigged configurations.   We in this paper   study  this correspondence by a comparison between the sets of the  rigged configurations  in the co-lexicographical order  and the  corresponding  sets of  the  Takahashi quantum numbers in  the lexicographical order.   As an example we consider the   $N=12$ case, because  the  solutions in this case  starts to  exhibit  several   interesting   structures which are not present  for any even length chain  for $N < 12$.  For example,   there are    solutions  in which some of the  individual   odd-length strings are not self-conjugate.    These solutions, known  non self-conjugate string solutions, have been observed  in \cite{giri} recently.   Another  type  of solutions, we call {\it almost singular}  string solutions,  also  start to  appear from  $N=12$,  in which  one pair of the rapidities takes the form of eq.  (\ref{asingular})  with $\epsilon^N \neq 0$.   The existence of   $\epsilon^N \neq 0$  is very much important theoretically   in the   calculations of  the  form   factors, scalar product  of the Bethe states. 
We obtained the  Bethe quantum numbers and the Takahashi quantum numbers, which are essential   for the purpose of  obtaining a correspondence  between the solutions of the Bethe equations and the rigged configurations. 

\begin{table}
\renewcommand\thetable{1}
\caption{Bethe roots for $N=12$, $M=1$. There are  $11$  solutions with one $1$-string}
\begin{tabular}{l@{\hskip 0.2in}r@{\hskip 0.2in}r@{\hskip 0.2in}l@{\hskip 0.2in}l@{\hskip 0.2in}l@{\hskip 0.2in}}
\hline\hline
&$J$& $I_n$ & \hspace{0.5in} $\lambda$& \hspace{0.5in} $E$ & Rigged Configuration\\
\hline
\multirow{1}*{1.}&\num{0}&\multirow{1}*{$\num{0}_1$} & \num{0} & \multirow{1}*{\num{ -2}} &
\multirow{1}*{\ytableausetup{centertableaux}
\begin{ytableau}
\none[10]&   &\none[5] 
\end{ytableau}}\\
&&&&&\\

\multirow{1}*{2.}&\num{-1}&\multirow{1}*{$\num{-1}_1$} & \num{-0.1339745962155614} & \multirow{1}*{\num{-1.8660254037844388}} &
\multirow{1}*{\ytableausetup{centertableaux}
\begin{ytableau}
\none[10]&   &\none[4] 
\end{ytableau}}\\
&&&&&\\
\multirow{1}*{3.}&\num{1}&\multirow{1}*{$\num{1}_1$} & \num{0.1339745962155614} & \multirow{1}*{\num{-1.8660254037844388}} &
\multirow{1}*{\ytableausetup{centertableaux}
\begin{ytableau}
\none[10]&   &\none[6] 
\end{ytableau}}\\
&&&&&\\
\multirow{1}*{4.}&\num{-2}&\multirow{1}*{$\num{-2}_1$} & \num{-0.2886751345948129} & \multirow{1}*{\num{ -1.4999999999999998}} &
\multirow{1}*{\ytableausetup{centertableaux}
\begin{ytableau}
\none[10]&   &\none[3] 
\end{ytableau}}\\
\hline
\end{tabular}
\end{table}

\begin{table}
 \renewcommand\thetable{2.1}
\caption{Bethe roots    for $N=12$, $M=2$. There are  $45$ solutions with two  $1$-strings}
\begin{tabular}{l@{\hskip 0.2in}r@{\hskip 0.2in}r@{\hskip 0.2in}l@{\hskip 0.2in}l@{\hskip 0.2in}l@{\hskip 0.2in}}
\hline\hline
&$J$& $I_n$ & $\lambda$& \hspace{0.1in} $E$ & Rigged Configuration\\
\hline
\multirow{2}*{1.}&\num{-1/2}&\multirow{1}*{$\num{-1/2}_1$} & \num{-0.07188914699748664} & \multirow{2}*{\num{-3.918985947228995}} &
\multirow{2}*{\ytableausetup{centertableaux}
\begin{ytableau}
\none[8]&   &\none[4] \\
\none[8]&   &\none[4]
\end{ytableau}}\\
&\num{1/2}&\multirow{1}*{$\num{1/2}_1$} &\num{ 0.07188914699748664} & &\\
&&&&&\\

\multirow{2}*{2.}&\num{-1/2}&\multirow{1}*{$\num{-1/2}_1$} & \num{-0.05930327105788041} & \multirow{2}*{\num{-3.660965635573501}} &
\multirow{2}*{\ytableausetup{centertableaux}
\begin{ytableau}
\none[8]&   &\none[3] \\
\none[8]&   &\none[3]
\end{ytableau}}\\
&\num{-3/2}&\multirow{1}*{$\num{-3/2}_1$} &\num{ -0.21467167951993998} & &\\
&&&&&\\

\multirow{2}*{3.}&\num{1/2}&\multirow{1}*{$\num{1/2}_1$} & \num{0.05930327105788041} & \multirow{2}*{\num{-3.660965635573501}} &
\multirow{2}*{\ytableausetup{centertableaux}
\begin{ytableau}
\none[8]&   &\none[5] \\
\none[8]&   &\none[5]
\end{ytableau}}\\
\hline
\end{tabular}
\end{table}

\begin{table}
 \renewcommand\thetable{2.1}
\caption{Continued}
\begin{tabular}{l@{\hskip 0.2in}r@{\hskip 0.2in}r@{\hskip 0.2in}l@{\hskip 0.2in}l@{\hskip 0.2in}l@{\hskip 0.2in}}
\hline\hline
&$J$& $I_n$ & $\lambda$& \hspace{0.1in} $E$ & Rigged Configuration\\
\hline

\multirow{2}*{22.}&\num{-3/2}&\multirow{1}*{$\num{-3/2}_1$} & \num{-0.24642588307230243} & \multirow{2}*{\num{ -2.2268893359696103}} &
\multirow{2}*{\ytableausetup{centertableaux}
\begin{ytableau}
\none[8]&   &\none[3] \\
\none[8]&   &\none[7]
\end{ytableau}}\\
&\num{7/2}&\multirow{1}*{$\num{7/2}_1$} &\num{ 0.7479203249834772} & &\\
&&&&&\\

\multirow{2}*{23.}&\num{3/2}&\multirow{1}*{$\num{3/2}_1$} & \num{0.24642588307230243} & \multirow{2}*{\num{ -2.2268893359696103}} &
\multirow{2}*{\ytableausetup{centertableaux}
\begin{ytableau}
\none[8]&   &\none[1] \\
\none[8]&   &\none[5]
\end{ytableau}}\\
&\num{-7/2}&\multirow{1}*{$\num{-7/2}_1$} &\num{ -0.7479203249834772} & &\\
&&&&&\\ 

\multirow{2}*{24.}&\num{-1/2}&\multirow{1}*{$\num{-1/2}_1$} & \num{-0.02404403589757235} & \multirow{2}*{\num{ 
-2.1813335297465004}} &
\multirow{2}*{\ytableausetup{centertableaux}
\begin{ytableau}
\none[8]&   &\none[0] \\
\none[8]&   &\none[3]
\end{ytableau}}\\
\hline
\end{tabular}
\end{table}

\begin{table}
 \renewcommand\thetable{2.1}
\caption{Continued}
\begin{tabular}{l@{\hskip 0.2in}r@{\hskip 0.2in}r@{\hskip 0.2in}l@{\hskip 0.2in}l@{\hskip 0.2in}l@{\hskip 0.2in}}
\hline\hline
&$J$& $I_n$ & $\lambda$& \hspace{0.1in} $E$ & Rigged Configuration\\
\hline

\multirow{2}*{43.}&\num{-7/2}&\multirow{1}*{$\num{-7/2}_1$} & \num{ -0.8049726473325098} & \multirow{2}*{\num{ -0.7247737335169068}} &
\multirow{2}*{\ytableausetup{centertableaux}
\begin{ytableau}
\none[8]&   &\none[1] \\
\none[8]&   &\none[8]
\end{ytableau}}\\
&\num{9/2}&\multirow{1}*{$\num{9/2}_1$}&\num{ 1.6512839711504512} & &\\
&&&&&\\

\multirow{2}*{44.}&\num{7/2}&\multirow{1}*{$\num{7/2}_1$} & \num{ 0.8049726473325098} & \multirow{2}*{\num{ -0.7247737335169068}} &
\multirow{2}*{\ytableausetup{centertableaux}
\begin{ytableau}
\none[8]&   &\none[0] \\
\none[8]&   &\none[7]
\end{ytableau}}\\
&\num{-9/2}&\multirow{1}*{$\num{-9/2}_1$}&\num{ -1.6512839711504512} & &\\
&&&&&\\

\multirow{2}*{45.}&\num{-9/2}&\multirow{1}*{$\num{-9/2}_1$} & \num{-1.7028436194446244} & \multirow{2}*{\num{ -0.31749293433763787}} &
\multirow{2}*{\ytableausetup{centertableaux}
\begin{ytableau}
\none[8]&   &\none[0] \\
\none[8]&   &\none[8]
\end{ytableau}}\\
&\num{9/2}&\multirow{1}*{$\num{9/2}_1$}&\num{ 1.7028436194446244} & &\\
&&&&&\\
\hline
\end{tabular}
\end{table}


\begin{table}
 \renewcommand\thetable{2.2}
\caption{Bethe roots for  $N=12, M= 2$. There are $9$ solutions with one $2$-string}
\begin{tabular}{l@{\hskip 0.2in}r@{\hskip 0.2in}r@{\hskip 0.2in}l@{\hskip 0.2in}l@{\hskip 0.2in}l@{\hskip 0.2in}}
\hline\hline
&$J$& $I_n$ & $\lambda$& \hspace{0.1in} $E$ & Rigged Configuration\\
\hline

\multirow{2}*{46.}&\num{5/2}&\multirow{2}*{$\num{0}_2$} & \num{ 0.5 i} & \multirow{2}*{\num{ -1}} &
\multirow{2}*{\ytableausetup{centertableaux}
\begin{ytableau}
\none[8]&  & &\none[4] 
\end{ytableau}}\\
&\num{7/2}&&\num{-0.5 i} & &\\
&&&&&\\

\multirow{2}*{47.}&\num{-7/2}&\multirow{2}*{$\num{-1}_2$} & \num{ -0.26794952964718094 + 0.49999990964310753 i} & \multirow{2}*{\num{ -0.9330138760941911}} &
\multirow{2}*{\ytableausetup{centertableaux}
\begin{ytableau}
\none[8]&  & &\none[3] 
\end{ytableau}}\\
&\num{-7/2}&&\num{-0.26794952964718094 - 0.49999990964310753  i} & &\\
&&&&&\\

\multirow{2}*{48.}&\num{7/2}&\multirow{2}*{$\num{1}_2$} & \num{ 0.26794952964718094 + 0.49999990964310753 i} & \multirow{2}*{\num{ -0.9330138760941911}} &
\multirow{2}*{\ytableausetup{centertableaux}
\begin{ytableau}
\none[8]&  & &\none[5] 
\end{ytableau}}\\
\hline
\end{tabular}
\end{table}

\begin{table}
 \renewcommand\thetable{3.1}
\caption{Bethe roots for $N=12$, $M=3$. There are  $84$ solutions with three  $1$-strings}
\begin{tabular}{l@{\hskip 0.2in}r@{\hskip 0.2in}r@{\hskip 0.2in}l@{\hskip 0.2in}l@{\hskip 0.2in}l@{\hskip 0.2in}}
\hline\hline
&$J$& $I_n$ & $\lambda$& \hspace{0.1in} $E$ & Rigged Configuration\\
\hline
\multirow{3}*{1.}&\num{-1}&\multirow{1}*{$\num{-1}_1$} & \num{ -0.1544087600908072} & \multirow{3}*{\num{ -5.65173991547287}} &
\multirow{3}*{\ytableausetup{centertableaux}
\begin{ytableau}
\none[6]&   &\none[3] \\
\none[6]&   &\none[3]\\
\none[6]&   &\none[3]
\end{ytableau}}\\
&\num{0}&\multirow{1}*{$\num{0}_1$} &\num{ 0} & &\\
&\num{1}&\multirow{1}*{$\num{1}_1$} &\num{ 0.1544087600908072} & &\\
&&&&&\\
&&&&&\\

\multirow{3}*{2.}&\num{0}&\multirow{1}*{$\num{0}_1$} & \num{ -0.019876862739478456} & \multirow{3}*{\num{ -5.2903940178213364}} &
\multirow{3}*{\ytableausetup{centertableaux}
\begin{ytableau}
\none[6]&   &\none[4] \\
\none[6]&   &\none[4]\\
\none[6]&   &\none[4]
\end{ytableau}}\\
&\num{1}&\multirow{1}*{$\num{1}_1$} &\num{ 0.13255745287590923} & &\\
\hline
\end{tabular}
\end{table}

\begin{table}
 \renewcommand\thetable{3.1}
\caption{Continued}
\begin{tabular}{l@{\hskip 0.2in}r@{\hskip 0.2in}r@{\hskip 0.2in}l@{\hskip 0.2in}l@{\hskip 0.2in}l@{\hskip 0.2in}}
\hline\hline
&$J$& $I_n$ & $\lambda$& \hspace{0.1in} $E$ & Rigged Configuration\\
\hline

\multirow{3}*{13.}&\num{-2}&\multirow{1}*{$\num{-2}_1$} & \num{-0.3341449486771437} & \multirow{3}*{\num{ -4.564315266954211}} &
\multirow{3}*{\ytableausetup{centertableaux}
\begin{ytableau}
\none[6]&   &\none[2] \\
\none[6]&   &\none[2]\\
\none[6]&   &\none[4]
\end{ytableau}}\\
&\num{-1}&\multirow{1}*{$\num{-1}_1$} &\num{ -0.14637777964821047} & &\\
&\num{2}&\multirow{1}*{$\num{2}_1$} &\num{ 0.3510421751176647} & &\\
&&&&&\\
&&&&&\\

\multirow{3}*{14.}&\num{2}&\multirow{1}*{$\num{2}_1$} & \num{0.3341449486771437} & \multirow{3}*{\num{ -4.564315266954211}} &
\multirow{3}*{\ytableausetup{centertableaux}
\begin{ytableau}
\none[6]&   &\none[2] \\
\none[6]&   &\none[4]\\
\none[6]&   &\none[4]
\end{ytableau}}\\
&\num{1}&\multirow{1}*{$\num{1}_1$} &\num{ 0.14637777964821047} & &\\
\hline
\end{tabular}
\end{table}

\begin{table}
 \renewcommand\thetable{3.1}
\caption{Continued}
\begin{tabular}{l@{\hskip 0.2in}r@{\hskip 0.2in}r@{\hskip 0.2in}l@{\hskip 0.2in}l@{\hskip 0.2in}l@{\hskip 0.2in}}
\hline\hline
&$J$& $I_n$ & $\lambda$& \hspace{0.1in} $E$ & Rigged Configuration\\
\hline

\multirow{3}*{26.}&\num{3}&\multirow{1}*{$\num{3}_1$} & \num{ 0.5846620783403371} & \multirow{3}*{\num{ -4.079685034143173}} &
\multirow{3}*{\ytableausetup{centertableaux}
\begin{ytableau}
\none[6]&   &\none[3] \\
\none[6]&   &\none[5]\\
\none[6]&   &\none[5]
\end{ytableau}}\\
&\num{2}&\multirow{1}*{$\num{2}_1$} &\num{ 0.29825597487982364} & &\\
&\num{-1}&\multirow{1}*{$\num{-1}_1$} &\num{-0.1847541524885523} & &\\
&&&&&\\
&&&&&\\

\multirow{3}*{27.}&\num{-4}&\multirow{1}*{$\num{-4}_1$} & \num{ -1.3118278812873365} & \multirow{3}*{\num{-4.001880584025997}} &
\multirow{3}*{\ytableausetup{centertableaux}
\begin{ytableau}
\none[6]&   &\none[0] \\
\none[6]&   &\none[3]\\
\none[6]&   &\none[3]
\end{ytableau}}\\
&\num{0}&\multirow{1}*{$\num{0}_1$} &\num{ 0.03249555441334122} & &\\
\hline
\end{tabular}
\end{table}

\begin{table}
 \renewcommand\thetable{3.1}
\caption{Continued}
\begin{tabular}{l@{\hskip 0.2in}r@{\hskip 0.2in}r@{\hskip 0.2in}l@{\hskip 0.2in}l@{\hskip 0.2in}l@{\hskip 0.2in}}
\hline\hline
&$J$& $I_n$ & $\lambda$& \hspace{0.1in} $E$ & Rigged Configuration\\
\hline

\multirow{3}*{39.}&\num{-3}&\multirow{1}*{$\num{-3}_1$} & \num{-0.5962010658484919} & \multirow{3}*{\num{-3.5721706585571034}} &
\multirow{3}*{\ytableausetup{centertableaux}
\begin{ytableau}
\none[6]&   &\none[1] \\
\none[6]&   &\none[1]\\
\none[6]&   &\none[4]
\end{ytableau}}\\
&\num{-2}&\multirow{1}*{$\num{-2}_1$} &\num{-0.30630796118398046} & &\\
&\num{2}&\multirow{1}*{$\num{2}_1$} &\num{0.3700838210822063} & &\\
&&&&&\\
&&&&&\\

\multirow{3}*{40.}&\num{3}&\multirow{1}*{$\num{3}_1$} & \num{ 0.5962010658484919} & \multirow{3}*{\num{-3.5721706585571034}} &
\multirow{3}*{\ytableausetup{centertableaux}
\begin{ytableau}
\none[6]&   &\none[2] \\
\none[6]&   &\none[5]\\
\none[6]&   &\none[5]
\end{ytableau}}\\
&\num{ 2}&\multirow{1}*{$\num{2}_1$} &\num{ 0.30630796118398046} & &\\
\hline
\end{tabular}
\end{table}

\begin{table}
 \renewcommand\thetable{3.1}
\caption{Continued}
\begin{tabular}{l@{\hskip 0.2in}r@{\hskip 0.2in}r@{\hskip 0.2in}l@{\hskip 0.2in}l@{\hskip 0.2in}l@{\hskip 0.2in}}
\hline\hline
&$J$& $I_n$ & $\lambda$& \hspace{0.1in} $E$ & Rigged Configuration\\
\hline

\multirow{3}*{52.}&\num{-4}&\multirow{1}*{$\num{-4}_1$} & \num{-1.1675150769433598} & \multirow{3}*{\num{ -3.295099959901812}} &
\multirow{3}*{\ytableausetup{centertableaux}
\begin{ytableau}
\none[6]&   &\none[0] \\
\none[6]&   &\none[0]\\
\none[6]&   &\none[1]
\end{ytableau}}\\
&\num{-3}&\multirow{1}*{$\num{-3}_1$} &\num{ -0.482011524202503} & &\\
&\num{-1}&\multirow{1}*{$\num{-1}_1$} &\num{ -0.08127735186955481} & &\\
&&&&&\\
&&&&&\\

\multirow{3}*{53.}&\num{4}&\multirow{1}*{$\num{4}_1$} & \num{1.1675150769433598} & \multirow{3}*{\num{ -3.295099959901812}} &
\multirow{3}*{\ytableausetup{centertableaux}
\begin{ytableau}
\none[6]&   &\none[5] \\
\none[6]&   &\none[6]\\
\none[6]&   &\none[6]
\end{ytableau}}\\
&\num{3}&\multirow{1}*{$\num{3}_1$} &\num{ 0.482011524202503} & &\\
\hline
\end{tabular}
\end{table}

\begin{table}
 \renewcommand\thetable{3.1}
\caption{Continued}
\begin{tabular}{l@{\hskip 0.2in}r@{\hskip 0.2in}r@{\hskip 0.2in}l@{\hskip 0.2in}l@{\hskip 0.2in}l@{\hskip 0.2in}}
\hline\hline
&$J$& $I_n$ & $\lambda$& \hspace{0.1in} $E$ & Rigged Configuration\\
\hline

\multirow{3}*{65.}&\num{3}&\multirow{1}*{$\num{3}_1$} & \num{ 0.610660152331662} & \multirow{3}*{\num{ -2.9538667921454884}} &
\multirow{3}*{\ytableausetup{centertableaux}
\begin{ytableau}
\none[6]&   &\none[1] \\
\none[6]&   &\none[5]\\
\none[6]&   &\none[5]
\end{ytableau}}\\
&\num{2}&\multirow{1}*{$\num{2}_1$} &\num{ 0.3165488028469942} & &\\
&\num{-3}&\multirow{1}*{$\num{-3}_1$} &\num{ -0.6641894231367725} & &\\
&&&&&\\
&&&&&\\

\multirow{3}*{66.}&\num{-4}&\multirow{1}*{$\num{-4}_1$} & \num{ -1.3378796871051393} & \multirow{3}*{\num{ -2.8345851482356275}} &
\multirow{3}*{\ytableausetup{centertableaux}
\begin{ytableau}
\none[6]&   &\none[0] \\
\none[6]&   &\none[2]\\
\none[6]&   &\none[5]
\end{ytableau}}\\
&\num{-1}&\multirow{1}*{$\num{-1}_1$} &\num{ -0.12481883420723379} & &\\
\hline
\end{tabular}
\end{table}

\begin{table}
 \renewcommand\thetable{3.1}
\caption{Continued}
\begin{tabular}{l@{\hskip 0.2in}r@{\hskip 0.2in}r@{\hskip 0.2in}l@{\hskip 0.2in}l@{\hskip 0.2in}l@{\hskip 0.2in}}
\hline\hline
&$J$& $I_n$ & $\lambda$& \hspace{0.1in} $E$ & Rigged Configuration\\
\hline

\multirow{3}*{77.}&\num{-4}&\multirow{1}*{$\num{-4}_1$} & \num{-1.3839135443853003} & \multirow{3}*{\num{ -2.307486473704967}} &
\multirow{3}*{\ytableausetup{centertableaux}
\begin{ytableau}
\none[6]&   &\none[0] \\
\none[6]&   &\none[2]\\
\none[6]&   &\none[6]
\end{ytableau}}\\
&\num{ -1}&\multirow{1}*{$\num{-1}_1$} &\num{-0.13877808723547794} & &\\
&\num{ 4}&\multirow{1}*{$\num{4}_1$} &\num{1.4236188036031703} & &\\
&&&&&\\
&&&&&\\

\multirow{3}*{78.}&\num{4}&\multirow{1}*{$\num{4}_1$} & \num{1.3839135443853003} & \multirow{3}*{\num{ -2.307486473704967}} &
\multirow{3}*{\ytableausetup{centertableaux}
\begin{ytableau}
\none[6]&   &\none[0] \\
\none[6]&   &\none[4]\\
\none[6]&   &\none[6]
\end{ytableau}}\\
&\num{ 1}&\multirow{1}*{$\num{1}_1$} &\num{ 0.13877808723547794} & &\\
\hline
\end{tabular}
\end{table}

\begin{table}
 \renewcommand\thetable{3.2}
\caption{Bethe  roots  for $N=12$, $M=3$. There are $63$ solutions with  one $1$-string  and one  $2$-string}
\begin{tabular}{l@{\hskip 0.2in}r@{\hskip 0.2in}r@{\hskip 0.2in}l@{\hskip 0.2in}l@{\hskip 0.2in}l@{\hskip 0.2in}}
\hline\hline
&$J$ & $I_n$& $\lambda$&\hspace{0.1in} $E$ & Rigged Configuration\\
\hline

\multirow{3}*{85.}&\num{0} & $\num{0}_1$& \num{0} & \multirow{3}*{\num{-3.0}}&
\multirow{3}*{\ytableausetup{centertableaux}
\begin{ytableau}
\none[6]&  & &\none[3] \\
\none[8] & &\none[4] 
\end{ytableau}}\\
&\num{3} &\multirow{2}*{$\num{0}_2$}&\num{ 0.5i} & &\\
&\num{3}& &\num{-0.5i}& &\\
&&&&&\\

\multirow{3}*{86.}&\num{0} & $\num{0}_1$& \num{0.03827169212800727} & \multirow{3}*{\num{-2.8782387358391652}}&
\multirow{3}*{\ytableausetup{centertableaux}
\begin{ytableau}
\none[6]&  & &\none[2] \\
\none[8] & &\none[4] 
\end{ytableau}}\\
&\num{-4} &\multirow{2}*{$\num{-1}_2$}&\num{ -0.3516936128362382 + 0.5000043473471327 i} & &\\
&\num{-3}& &\num{-0.3516936128362382 - 0.5000043473471327 i}& &\\
\hline
\end{tabular}
\end{table}

\begin{table}
 \renewcommand\thetable{3.2}
\caption{(Continued)}
\begin{tabular}{l@{\hskip 0.2in}r@{\hskip 0.2in}r@{\hskip 0.2in}l@{\hskip 0.2in}l@{\hskip 0.2in}l@{\hskip 0.2in}}
\hline\hline
&$J$ & $I_n$& $\lambda$&\hspace{0.1in} $E$ & Rigged Configuration\\
\hline

\multirow{3}*{100.}&\num{2} & $\num{2}_1$& \num{0.346408430776481} & \multirow{3}*{\num{-2.344599357264179}}&
\multirow{3}*{\ytableausetup{centertableaux}
\begin{ytableau}
\none[6]&  & &\none[3] \\
\none[8] & &\none[6] 
\end{ytableau}}\\
&\num{-3} &\multirow{2}*{$\num{0}_2$}&\num{ -0.08247625513898412 + 0.49999999999977472 i} & &\\
&\num{-3}& &\num{-0.08247625513898412 - 0.49999999999977472 i}& &\\
&&&&&\\

\multirow{3}*{101.}&\num{-2} & $\num{-2}_1$& \num{-0.346408430776481} & \multirow{3}*{\num{-2.344599357264179}}&
\multirow{3}*{\ytableausetup{centertableaux}
\begin{ytableau}
\none[6]&  & &\none[3] \\
\none[8] & &\none[2] 
\end{ytableau}}\\
&\num{3} &\multirow{2}*{$\num{0}_2$}&\num{ 0.08247625513898412 + 0.49999999999977472 i} & &\\
&\num{3}& &\num{0.08247625513898412 - 0.49999999999977472 i}& &\\
\hline
\end{tabular}
\end{table}

\begin{table}
 \renewcommand\thetable{3.2}
\caption{(Continued)}
\begin{tabular}{l@{\hskip 0.2in}r@{\hskip 0.2in}r@{\hskip 0.2in}l@{\hskip 0.2in}l@{\hskip 0.2in}l@{\hskip 0.2in}}
\hline\hline
&$J$ & $I_n$& $\lambda$&\hspace{0.1in} $E$ & Rigged Configuration\\
\hline

\multirow{3}*{116.}&\num{-2} & $\num{-2}_1$& \num{-0.40319480804654456} & \multirow{3}*{\num{-1.8234150347240905}}&
\multirow{3}*{\ytableausetup{centertableaux}
\begin{ytableau}
\none[6]&  & &\none[5] \\
\none[8] & &\none[2] 
\end{ytableau}}\\
&\num{ 4} &\multirow{2}*{$\num{2}_2$}&\num{ 0.8138087664957081 + 0.49392532873137657 i} & &\\
&\num{ 4}& &\num{0.8138087664957081 - 0.49392532873137657 i}& &\\
&&&&&\\

\multirow{3}*{117.}&\num{ 2} & $\num{2}_1$& \num{ 0.40319480804654456} & \multirow{3}*{\num{-1.8234150347240905}}&
\multirow{3}*{\ytableausetup{centertableaux}
\begin{ytableau}
\none[6]&  & &\none[1] \\
\none[8] & &\none[6] 
\end{ytableau}}\\
&\num{ -4} &\multirow{2}*{$\num{-2}_2$}&\num{ -0.8138087664957081 + 0.49392532873137657 i} & &\\
&\num{ -4}& &\num{-0.8138087664957081 - 0.49392532873137657 i}& &\\
\hline
\end{tabular}
\end{table}

\begin{table}
 \renewcommand\thetable{3.2}
\caption{(Continued)}
\begin{tabular}{l@{\hskip 0.2in}r@{\hskip 0.2in}r@{\hskip 0.2in}l@{\hskip 0.2in}l@{\hskip 0.2in}l@{\hskip 0.2in}}
\hline\hline
&$J$ & $I_n$& $\lambda$&\hspace{0.1in} $E$ & Rigged Configuration\\
\hline

\multirow{3}*{132.}&\num{4} & $\num{4}_1$& \num{1.2863097990446577} & \multirow{3}*{\num{-1.250703216466408}}&
\multirow{3}*{\ytableausetup{centertableaux}
\begin{ytableau}
\none[6]&  & &\none[4] \\
\none[8] & &\none[8] 
\end{ytableau}}\\
&\num{ 3} &\multirow{2}*{$\num{1}_2$}&\num{0.10936846774865674 + 0.5000000000040663   i} & &\\
&\num{ 4}& &\num{0.10936846774865674 - 0.5000000000040663   i}& &\\
&&&&&\\

\multirow{3}*{133.}&\num{-4} & $\num{-4}_1$& \num{-1.2863097990446577} & \multirow{3}*{\num{-1.250703216466408}}&
\multirow{3}*{\ytableausetup{centertableaux}
\begin{ytableau}
\none[6]&  & &\none[2] \\
\none[8] & &\none[0] 
\end{ytableau}}\\
&\num{ -4} &\multirow{2}*{$\num{-1}_2$}&\num{-0.10936846774865674 + 0.5000000000040663   i} & &\\
&\num{ -3}& &\num{-0.10936846774865674 - 0.5000000000040663   i}& &\\
\hline
\end{tabular}
\end{table}

\begin{table}
 \renewcommand\thetable{3.3}
\caption{Bethe roots  for $N=12$, $M=3$. There are $7$ solutions with  one $3$-string}
\begin{tabular}{l@{\hskip 0.2in}r@{\hskip 0.2in}r@{\hskip 0.2in}l@{\hskip 0.2in}l@{\hskip 0.2in}l@{\hskip 0.2in}}
\hline\hline
&$J$ & $I_n$& $\lambda$&\hspace{0.1in} $E$ & Rigged Configuration\\
\hline
\multirow{3}*{148.}&\num{5} & \multirow{3}*{$\num{0}_3$}& \num{+1.0000112919922328 i} & \multirow{3}*{\num{-0.666706814990114}}&
\multirow{3}*{\ytableausetup{centertableaux}
\begin{ytableau}
\none[6]&  & & &\none[3] 
\end{ytableau}}\\
&\num{ 0} & &\num{0} & &\\
&\num{ -5}& &\num{-1.0000112919922328   i}& &\\
&&&&&\\

\multirow{3}*{149.}&\num{-6} & \multirow{3}*{$\num{-1}_3$}& \num{-0.40159436544190885 + 1.0000391604261667 i} & \multirow{3}*{\num{-0.6217449658211609}}&
\multirow{3}*{\ytableausetup{centertableaux}
\begin{ytableau}
\none[6]&  & & &\none[2] 
\end{ytableau}}\\
&\num{ -2} & &\num{-0.40173685251462715} & &\\
&\num{ -5}& &\num{-0.40159436544190885 - 1.0000391604261667  i}& &\\
\hline
\end{tabular}
\end{table}

\begin{table}
 \renewcommand\thetable{4.1}
\caption{Bethe roots  for $N=12$, $M=4$. There are  $70$ solutions with four  $1$-strings}
\begin{tabular}{l@{\hskip 0.2in}r@{\hskip 0.2in}r@{\hskip 0.2in}l@{\hskip 0.2in}l@{\hskip 0.2in}l@{\hskip 0.2in}}
\hline\hline
&$J$& $I_n$ & $\lambda$& \hspace{0.1in} $E$ & Rigged Configuration\\
\hline
\multirow{4}*{1.}&\num{-3/2}&\multirow{1}*{$\num{-3/2}_1$} & \num{ -0.25636163808139817} & \multirow{4}*{\num{-7.070529325963522}} &
\multirow{4}*{\ytableausetup{centertableaux}
\begin{ytableau}
\none[4]&   &\none[2] \\
\none[4]&   &\none[2]\\
\none[4]&   &\none[2]\\
\none[4]&   &\none[2]
\end{ytableau}}\\
&\num{-1/2}&\multirow{1}*{$\num{-1/2}_1$} &\num{-0.07874918499591191} & &\\
&\num{1/2}&\multirow{1}*{$\num{1/2}_1$} &\num{0.07874918499591191} & &\\
&\num{3/2}&\multirow{1}*{$\num{3/2}_1$} &\num{ 0.25636163808139817} & &\\
&&&&&\\
&&&&&\\
\multirow{4}*{2.}&\num{-5/2}&\multirow{1}*{$\num{-5/2}_1$} & \num{-0.46701763509344507} & \multirow{4}*{\num{-6.6374064251773826}} &
\multirow{4}*{\ytableausetup{centertableaux}
\begin{ytableau}
\none[4]&   &\none[1] \\
\none[4]&   &\none[1]\\
\none[4]&   &\none[1]\\
\none[4]&   &\none[1]
\end{ytableau}}\\
\hline
\end{tabular}
\end{table}

\begin{table}
 \renewcommand\thetable{4.1}
\caption{(Continued)}
\begin{tabular}{l@{\hskip 0.2in}r@{\hskip 0.2in}r@{\hskip 0.2in}l@{\hskip 0.2in}l@{\hskip 0.2in}l@{\hskip 0.2in}}
\hline\hline
&$J$& $I_n$ & $\lambda$& \hspace{0.1in} $E$ & Rigged Configuration\\
\hline

\multirow{4}*{11.}&\num{7/2}&\multirow{1}*{$\num{7/2}_1$} & \num{0.9992907814110428} & \multirow{4}*{\num{-6.0097169056126765}} &
\multirow{4}*{\ytableausetup{centertableaux}
\begin{ytableau}
\none[4]&   &\none[3] \\
\none[4]&   &\none[3]\\
\none[4]&   &\none[3]\\
\none[4]&   &\none[4]
\end{ytableau}}\\
&\num{ 3/2}&\multirow{1}*{$\num{3/2}_1$} &\num{ 0.19720687795494574} & &\\
&\num{1/2}&\multirow{1}*{$\num{1/2}_1$} &\num{ 0.03313610257555372} & &\\
&\num{-1/2}&\multirow{1}*{$\num{-1/2}_1$} &\num{-0.12221249974628126} & &\\
&&&&&\\
&&&&&\\

\multirow{4}*{12.}&\num{-5/2}&\multirow{1}*{$\num{-5/2}_1$} & \num{-0.5149274860657148} & \multirow{4}*{\num{-5.847128730476953}} &
\multirow{4}*{\ytableausetup{centertableaux}
\begin{ytableau}
\none[4]&   &\none[1] \\
\none[4]&   &\none[2]\\
\none[4]&   &\none[2]\\
\none[4]&   &\none[3]
\end{ytableau}}\\
\hline
\end{tabular}
\end{table}

\begin{table}
 \renewcommand\thetable{4.1}
\caption{(Continued) }
\begin{tabular}{l@{\hskip 0.2in}r@{\hskip 0.2in}r@{\hskip 0.2in}l@{\hskip 0.2in}l@{\hskip 0.2in}l@{\hskip 0.2in}}
\hline\hline
&$J$& $I_n$ & $\lambda$& \hspace{0.1in} $E$ & Rigged Configuration\\
\hline

\multirow{4}*{22.}&\num{7/2}&\multirow{1}*{$\num{7/2}_1$} & \num{0.9577990835437517} & \multirow{4}*{\num{-5.498295943848788}} &
\multirow{4}*{\ytableausetup{centertableaux}
\begin{ytableau}
\none[4]&   &\none[3] \\
\none[4]&   &\none[3]\\
\none[4]&   &\none[4]\\
\none[4]&   &\none[4]
\end{ytableau}}\\
&\num{ 5/2}&\multirow{1}*{$\num{5/2}_1$} &\num{0.4077488334999926} & &\\
&\num{1/2}&\multirow{1}*{$\num{1/2}_1$} &\num{0.02539719560259699} & &\\
&\num{-1/2}&\multirow{1}*{$\num{-1/2}_1$} &\num{-0.1296724445739804} & &\\
&&&&&\\
&&&&&\\

\multirow{4}*{23.}&\num{-7/2}&\multirow{1}*{$\num{-7/2}_1$} & \num{-0.9157444898681719} & \multirow{4}*{\num{-5.486250405908474}} &
\multirow{4}*{\ytableausetup{centertableaux}
\begin{ytableau}
\none[4]&   &\none[0] \\
\none[4]&   &\none[0]\\
\none[4]&   &\none[0]\\
\none[4]&   &\none[0]
\end{ytableau}}\\
\hline
\end{tabular}
\end{table}

\begin{table}
 \renewcommand\thetable{4.1}
\caption{(Continued) }
\begin{tabular}{l@{\hskip 0.2in}r@{\hskip 0.2in}r@{\hskip 0.2in}l@{\hskip 0.2in}l@{\hskip 0.2in}l@{\hskip 0.2in}}
\hline\hline
&$J$& $I_n$ & $\lambda$& \hspace{0.1in} $E$ & Rigged Configuration\\
\hline

\multirow{4}*{33.}&\num{7/2}&\multirow{1}*{$\num{7/2}_1$} & \num{0.9765133938142908} & \multirow{4}*{\num{-5.04567427020521}} &
\multirow{4}*{\ytableausetup{centertableaux}
\begin{ytableau}
\none[4]&   &\none[2] \\
\none[4]&   &\none[3]\\
\none[4]&   &\none[4]\\
\none[4]&   &\none[4]
\end{ytableau}}\\
&\num{ 5/2}&\multirow{1}*{$\num{5/2}_1$} &\num{0.4165665466287065} & &\\
&\num{1/2}&\multirow{1}*{$\num{1/2}_1$} &\num{0.032220801507225505} & &\\
&\num{-3/2}&\multirow{1}*{$\num{-3/2}_1$} &\num{-0.3048698222864835} & &\\
&&&&&\\
&&&&&\\

\multirow{4}*{34.}&\num{-7/2}&\multirow{1}*{$\num{-7/2}_1$} & \num{-0.9954862676993101} & \multirow{4}*{\num{-4.9409095270536}} &
\multirow{4}*{\ytableausetup{centertableaux}
\begin{ytableau}
\none[4]&   &\none[0] \\
\none[4]&   &\none[0]\\
\none[4]&   &\none[2]\\
\none[4]&   &\none[2]
\end{ytableau}}\\
\hline
\end{tabular}
\end{table}

\begin{table}
 \renewcommand\thetable{4.1}
\caption{(Continued) }
\begin{tabular}{l@{\hskip 0.2in}r@{\hskip 0.2in}r@{\hskip 0.2in}l@{\hskip 0.2in}l@{\hskip 0.2in}l@{\hskip 0.2in}}
\hline\hline
&$J$& $I_n$ & $\lambda$& \hspace{0.1in} $E$ & Rigged Configuration\\
\hline

\multirow{4}*{44.}&\num{-7/2}&\multirow{1}*{$\num{-7/2}_1$} & \num{-0.9528669544299918} & \multirow{4}*{\num{-4.827340110342111}} &
\multirow{4}*{\ytableausetup{centertableaux}
\begin{ytableau}
\none[4]&   &\none[0] \\
\none[4]&   &\none[0]\\
\none[4]&   &\none[0]\\
\none[4]&   &\none[2]
\end{ytableau}}\\
&\num{ -5/2}&\multirow{1}*{$\num{-5/2}_1$} &\num{-0.40660954439829283} & &\\
&\num{-3/2}&\multirow{1}*{$\num{-3/2}_1$} &\num{-0.1875932286290853} & &\\
&\num{3/2}&\multirow{1}*{$\num{3/2}_1$} &\num{0.31239383445814} & &\\
&&&&&\\
&&&&&\\

\multirow{4}*{45.}&\num{7/2}&\multirow{1}*{$\num{7/2}_1$} & \num{0.9528669544299918} & \multirow{4}*{\num{-4.827340110342111}} &
\multirow{4}*{\ytableausetup{centertableaux}
\begin{ytableau}
\none[4]&   &\none[2] \\
\none[4]&   &\none[4]\\
\none[4]&   &\none[4]\\
\none[4]&   &\none[4]
\end{ytableau}}\\
\hline
\end{tabular}
\end{table}

\begin{table}
 \renewcommand\thetable{4.1}
\caption{(Continued) }
\begin{tabular}{l@{\hskip 0.2in}r@{\hskip 0.2in}r@{\hskip 0.2in}l@{\hskip 0.2in}l@{\hskip 0.2in}l@{\hskip 0.2in}}
\hline\hline
&$J$& $I_n$ & $\lambda$& \hspace{0.1in} $E$ & Rigged Configuration\\
\hline
\multirow{4}*{55.}&\num{-7/2}&\multirow{1}*{$\num{-7/2}_1$} & \num{-1.1443233446935346} & \multirow{4}*{\num{-4.253476227520393}} &
\multirow{4}*{\ytableausetup{centertableaux}
\begin{ytableau}
\none[4]&   &\none[0] \\
\none[4]&   &\none[2]\\
\none[4]&   &\none[3]\\
\none[4]&   &\none[4]
\end{ytableau}}\\
&\num{ -1/2}&\multirow{1}*{$\num{-1/2}_1$} &\num{-0.08102022242684594} & &\\
&\num{3/2}&\multirow{1}*{$\num{3/2}_1$} &\num{0.23369152484367983} & &\\
&\num{7/2}&\multirow{1}*{$\num{7/2}_1$} &\num{1.0997545048566437} & &\\
&&&&&\\
&&&&&\\

\multirow{4}*{56.}&\num{7/2}&\multirow{1}*{$\num{7/2}_1$} & \num{1.1443233446935346} & \multirow{4}*{\num{-4.253476227520393}} &
\multirow{4}*{\ytableausetup{centertableaux}
\begin{ytableau}
\none[4]&   &\none[0] \\
\none[4]&   &\none[1]\\
\none[4]&   &\none[2]\\
\none[4]&   &\none[4]
\end{ytableau}}\\
\hline
\end{tabular}
\end{table}

\begin{table}
 \renewcommand\thetable{4.1}
\caption{(Continued) }
\begin{tabular}{l@{\hskip 0.2in}r@{\hskip 0.2in}r@{\hskip 0.2in}l@{\hskip 0.2in}l@{\hskip 0.2in}l@{\hskip 0.2in}}
\hline\hline
&$J$& $I_n$ & $\lambda$& \hspace{0.1in} $E$ & Rigged Configuration\\
\hline
\multirow{4}*{66.}&\num{-7/2}&\multirow{1}*{$\num{-7/2}_1$} & \num{-1.2018347224491464} & \multirow{4}*{\num{-3.5294629436140594}} &
\multirow{4}*{\ytableausetup{centertableaux}
\begin{ytableau}
\none[4]&   &\none[0] \\
\none[4]&   &\none[4]\\
\none[4]&   &\none[4]\\
\none[4]&   &\none[4]
\end{ytableau}}\\
&\num{ 3/2}&\multirow{1}*{$\num{3/2}_1$} &\num{0.20997739400374454} & &\\
&\num{5/2}&\multirow{1}*{$\num{5/2}_1$} &\num{0.4335082150646594} & &\\
&\num{7/2}&\multirow{1}*{$\num{7/2}_1$} &\num{1.011902837954899} & &\\
&&&&&\\
&&&&&\\

\multirow{4}*{67.}&\num{7/2}&\multirow{1}*{$\num{7/2}_1$} & \num{1.2018347224491464} & \multirow{4}*{\num{-3.5294629436140594}} &
\multirow{4}*{\ytableausetup{centertableaux}
\begin{ytableau}
\none[4]&   &\none[0] \\
\none[4]&   &\none[0]\\
\none[4]&   &\none[0]\\
\none[4]&   &\none[4]
\end{ytableau}}\\
\hline
\end{tabular}
\end{table}

\begin{table}
 \renewcommand\thetable{4.2}
\caption{Bethe roots for $N=12$, $M=4$. There are  $140$ solutions with two $1$-strings and one $2$-string}
\begin{tabular}{l@{\hskip 0.2in}r@{\hskip 0.2in}r@{\hskip 0.2in}l@{\hskip 0.2in}l@{\hskip 0.2in}l@{\hskip 0.2in}}
\hline\hline
&$J$& $I_n$ & $\lambda$& \hspace{0.1in} $E$ & Rigged Configuration\\
\hline
\multirow{4}*{71.}&\num{-1/2} &$\num{-1/2}_1$& \num{-0.08199356634008453} & \multirow{4}*{\num{-4.895249800546962}}&
\multirow{4}*{\ytableausetup{centertableaux}
\begin{ytableau}
\none[4]&  & &\none[2] \\
\none[6] & &\none[3] \\
\none[6] & &\none[3]
\end{ytableau}}\\
&\num{1/2} &$\num{1/2}_1$& \num{0.08199356634008453} & &\\
&\num{5/2}&\multirow{2}*{$\num{0}_2$}&\num{ 0.5i} & &\\
&\num{7/2}&&\num{ -0.5i} & &\\
&&&&&\\

\multirow{4}*{72.}&\num{-3/2} &$\num{-3/2}_1$& \num{-0.26185517516265894} &\multirow{4}*{\num{-4.506357797368622 } }&
\multirow{4}*{\ytableausetup{centertableaux}
\begin{ytableau}
\none[4]&  & &\none[2] \\
\none[6] & &\none[2] \\
\none[6] & &\none[2]
\end{ytableau}}\\
&\num{-1/2}&$\num{-1/2}_1$ & \num{-0.0797015330423064} && \\
&\num{5/2}&\multirow{2}*{$\num{0}_2$}&\num{0.11744543525923092 + 0.5000000000440222  i} & &\\
&\num{7/2}&&\num{ 0.11744543525923092 - 0.5000000000440222  i} & &\\
&&&&&\\

\multirow{4}*{73.}&\num{3/2} &$\num{3/2}_1$& \num{0.26185517516265894} &\multirow{4}*{\num{-4.506357797368622 } }&
\multirow{4}*{\ytableausetup{centertableaux}
\begin{ytableau}
\none[4]&  & &\none[2] \\
\none[6] & &\none[4] \\
\none[6] & &\none[4]
\end{ytableau}}\\
&\num{1/2}&$\num{1/2}_1$ & \num{0.0797015330423064} && \\
&\num{-7/2}&\multirow{2}*{$\num{0}_2$}&\num{-0.11744543525923092 + 0.5000000000440222  i} & &\\
&\num{-5/2}&&\num{ -0.11744543525923092 - 0.5000000000440222  i} & &\\
&&&&&\\

\hline
\end{tabular}
\end{table}

\begin{table}
 \renewcommand\thetable{4.2}
\caption{(Continued)}
\begin{tabular}{l@{\hskip 0.2in}r@{\hskip 0.2in}r@{\hskip 0.2in}l@{\hskip 0.2in}l@{\hskip 0.2in}l@{\hskip 0.2in}}
\hline\hline
&$J$& $I_n$ & $\lambda$& \hspace{0.1in} $E$ & Rigged Configuration\\
\hline
\multirow{4}*{74.}&\num{-3/2} &$\num{-3/2}_1$& \num{-0.2644265515068217} &\multirow{4}*{\num{-4.506202917280661} }&
\multirow{4}*{\ytableausetup{centertableaux}
\begin{ytableau}
\none[4]&  & &\none[2] \\
\none[6] & &\none[2] \\
\none[6] & &\none[3]
\end{ytableau}}\\
&\num{1/2}&$\num{1/2}_1$ & \num{0.08244177274885205} && \\
&\num{5/2}&\multirow{2}*{$\num{0}_2$}&\num{ 0.0613268445095665 + 0.5000000000000213 i} & &\\
&\num{7/2}&&\num{ 0.0613268445095665 - 0.5000000000000213 i} & &\\
&&&&&\\
\multirow{4}*{75.}&\num{3/2} &$\num{3/2}_1$& \num{0.2644265515068217} &\multirow{4}*{\num{-4.506202917280661} }&
\multirow{4}*{\ytableausetup{centertableaux}
\begin{ytableau}
\none[4]&  & &\none[2] \\
\none[6] & &\none[3] \\
\none[6] & &\none[4]
\end{ytableau}}\\
&\num{-1/2}&$\num{-1/2}_1$ & \num{-0.08244177274885205} && \\
\hline
\end{tabular}
\end{table}

\clearpage

\begin{table}
 \renewcommand\thetable{4.2}
\caption{(Continued)}
\begin{tabular}{l@{\hskip 0.2in}r@{\hskip 0.2in}r@{\hskip 0.2in}l@{\hskip 0.2in}l@{\hskip 0.2in}l@{\hskip 0.2in}}
\hline\hline
&$J$& $I_n$ & $\lambda$& \hspace{0.1in} $E$ & Rigged Configuration\\
\hline
\multirow{4}*{87.}&\num{-7/2} &$\num{-7/2}_1$& \num{-1.1147016456120633} &\multirow{4}*{\num{-3.2357220454136635} }&
\multirow{4}*{\ytableausetup{centertableaux}
\begin{ytableau}
\none[4]&  & &\none[2] \\
\none[6] & &\none[0] \\
\none[6] & &\none[3]
\end{ytableau}}\\
&\num{1/2}&$\num{1/2}_1$ & \num{0.09355235179702635} && \\
&\num{5/2}&\multirow{2}*{$\num{0}_2$}&\num{0.1807137554671831 + 0.500000004227036   i} & &\\
&\num{7/2}&&\num{0.1807137554671831 - 0.500000004227036   i} & &\\
&&&&&\\

\multirow{4}*{88.}&\num{7/2} &$\num{7/2}_1$& \num{1.1147016456120633} &\multirow{4}*{\num{-3.2357220454136635} }&
\multirow{4}*{\ytableausetup{centertableaux}
\begin{ytableau}
\none[4]&  & &\none[2] \\
\none[6] & &\none[3] \\
\none[6] & &\none[6]
\end{ytableau}}\\
&\num{-1/2}&$\num{-1/2}_1$ & \num{-0.09355235179702635} && \\
\hline
\end{tabular}
\end{table}

\begin{table}
 \renewcommand\thetable{4.2}
\caption{(Continued)}
\begin{tabular}{l@{\hskip 0.2in}r@{\hskip 0.2in}r@{\hskip 0.2in}l@{\hskip 0.2in}l@{\hskip 0.2in}l@{\hskip 0.2in}}
\hline\hline
&$J$& $I_n$ & $\lambda$& \hspace{0.1in} $E$ & Rigged Configuration\\
\hline
\multirow{4}*{99.}&\num{-1/2} &$\num{-1/2}_1$& \num{-0.12802823043278505} &\multirow{4}*{\num{-4.677654794255555} }&
\multirow{4}*{\ytableausetup{centertableaux}
\begin{ytableau}
\none[4]&  & &\none[3] \\
\none[6] & &\none[3] \\
\none[6] & &\none[3]
\end{ytableau}}\\
&\num{1/2}&$\num{1/2}_1$ & \num{0.028410190328398373} && \\
&\num{7/2}&\multirow{2}*{$\num{1}_2$}&\num{0.49062964345360927 + 0.4997488635033444    i} & &\\
&\num{7/2}&&\num{0.49062964345360927 - 0.4997488635033444    i} & &\\
&&&&&\\
\multirow{4}*{100.}&\num{1/2} &$\num{1/2}_1$& \num{0.12802823043278505} &\multirow{4}*{\num{-4.677654794255555} }&
\multirow{4}*{\ytableausetup{centertableaux}
\begin{ytableau}
\none[4]&  & &\none[1] \\
\none[6] & &\none[3] \\
\none[6] & &\none[3]
\end{ytableau}}\\
&\num{-1/2}&$\num{-1/2}_1$ & \num{-0.028410190328398373} && \\
\hline
\end{tabular}
\end{table}

\begin{table}
 \renewcommand\thetable{4.2}
\caption{(Continued)}
\begin{tabular}{l@{\hskip 0.2in}r@{\hskip 0.2in}r@{\hskip 0.2in}l@{\hskip 0.2in}l@{\hskip 0.2in}l@{\hskip 0.2in}}
\hline\hline
&$J$& $I_n$ & $\lambda$& \hspace{0.1in} $E$ & Rigged Configuration\\
\hline
\multirow{4}*{111.}&\num{-3/2} &$\num{-3/2}_1$& \num{-0.31046480333690735} &\multirow{4}*{\num{-3.9930804378607165} }&
\multirow{4}*{\ytableausetup{centertableaux}
\begin{ytableau}
\none[4]&  & &\none[3] \\
\none[6] & &\none[2] \\
\none[6] & &\none[4]
\end{ytableau}}\\
&\num{3/2}&$\num{3/2}_1$ & \num{0.1962553518025457} && \\
&\num{7/2}&\multirow{2}*{$\num{1}_2$}&\num{0.4753811002741261 + 0.4998097798825905  i} & &\\
&\num{7/2}&&\num{0.4753811002741261 - 0.4998097798825905   i} & &\\
&&&&&\\

\multirow{4}*{112.}&\num{3/2} &$\num{3/2}_1$& \num{0.31046480333690735} &\multirow{4}*{\num{-3.9930804378607165} }&
\multirow{4}*{\ytableausetup{centertableaux}
\begin{ytableau}
\none[4]&  & &\none[1] \\
\none[6] & &\none[2] \\
\none[6] & &\none[4]
\end{ytableau}}\\
&\num{-3/2}&$\num{-3/2}_1$ & \num{-0.1962553518025457} && \\
\hline
\end{tabular}
\end{table}

\begin{table}
 \renewcommand\thetable{4.2}
\caption{(Continued)}
\begin{tabular}{l@{\hskip 0.2in}r@{\hskip 0.2in}r@{\hskip 0.2in}l@{\hskip 0.2in}l@{\hskip 0.2in}l@{\hskip 0.2in}}
\hline\hline
&$J$& $I_n$ & $\lambda$& \hspace{0.1in} $E$ & Rigged Configuration\\
\hline
\multirow{4}*{123.}&\num{-5/2} &$\num{-5/2}_1$& \num{-0.5772775967442026} &\multirow{4}*{\num{-3.369467618447652} }&
\multirow{4}*{\ytableausetup{centertableaux}
\begin{ytableau}
\none[4]&  & &\none[3] \\
\none[6] & &\none[1] \\
\none[6] & &\none[4]
\end{ytableau}}\\
&\num{3/2}&$\num{3/2}_1$ & \num{0.20255892542544784} && \\
&\num{7/2}&\multirow{2}*{$\num{1}_2$}&\num{0.5112346581045921 + 0.49968029255201185    i} & &\\
&\num{7/2}&&\num{0.5112346581045921 - 0.49968029255201185   i} & &\\
&&&&&\\

\multirow{4}*{124.}&\num{5/2} &$\num{5/2}_1$& \num{0.5772775967442026} &\multirow{4}*{\num{-3.369467618447652} }&
\multirow{4}*{\ytableausetup{centertableaux}
\begin{ytableau}
\none[4]&  & &\none[1] \\
\none[6] & &\none[2] \\
\none[6] & &\none[5]
\end{ytableau}}\\
&\num{-3/2}&$\num{-3/2}_1$ & \num{-0.20255892542544784} && \\
\hline
\end{tabular}
\end{table}

\begin{table}
 \renewcommand\thetable{4.2}
\caption{(Continued)}
\begin{tabular}{l@{\hskip 0.2in}r@{\hskip 0.2in}r@{\hskip 0.2in}l@{\hskip 0.2in}l@{\hskip 0.2in}l@{\hskip 0.2in}}
\hline\hline
&$J$& $I_n$ & $\lambda$& \hspace{0.1in} $E$ & Rigged Configuration\\
\hline
\multirow{4}*{135.}&\num{-7/2} &$\num{-7/2}_1$& \num{-1.166931539668738} &\multirow{4}*{\num{-2.9656727606015107} }&
\multirow{4}*{\ytableausetup{centertableaux}
\begin{ytableau}
\none[4]&  & &\none[3] \\
\none[6] & &\none[0] \\
\none[6] & &\none[2]
\end{ytableau}}\\
&\num{-1/2}&$\num{-1/2}_1$ & \num{-0.09099583855398737} && \\
&\num{7/2}&\multirow{2}*{$\num{1}_2$}&\num{0.629795574272442 + 0.49874138047740846   i} & &\\
&\num{7/2}&&\num{0.629795574272442 - 0.49874138047740846   i} & &\\
&&&&&\\

\multirow{4}*{136.}&\num{7/2} &$\num{7/2}_1$& \num{1.166931539668738} &\multirow{4}*{\num{-2.9656727606015107} }&
\multirow{4}*{\ytableausetup{centertableaux}
\begin{ytableau}
\none[4]&  & &\none[1] \\
\none[6] & &\none[4] \\
\none[6] & &\none[6]
\end{ytableau}}\\
&\num{1/2}&$\num{1/2}_1$ & \num{0.09099583855398737} && \\
\hline
\end{tabular}
\end{table}

\clearpage

\begin{table}
 \renewcommand\thetable{4.2}
\caption{(Continued)}
\begin{tabular}{l@{\hskip 0.2in}r@{\hskip 0.2in}r@{\hskip 0.2in}l@{\hskip 0.2in}l@{\hskip 0.2in}l@{\hskip 0.2in}}
\hline\hline
&$J$& $I_n$ & $\lambda$& \hspace{0.1in} $E$ & Rigged Configuration\\
\hline

\multirow{4}*{148.}&\num{7/2} &$\num{7/2}_1$& \num{1.2113881068580123} &\multirow{4}*{\num{-2.2078526482728202} }&
\multirow{4}*{\ytableausetup{centertableaux}
\begin{ytableau}
\none[4]&  & &\none[1] \\
\none[6] & &\none[1] \\
\none[6] & &\none[6]
\end{ytableau}}\\
&\num{-5/2}&$\num{-5/2}_1$ & \num{-0.45173411523691614} && \\
&\num{-7/2}&\multirow{2}*{$\num{-1}_2$}&\num{-0.47676177170605344 + 0.4998447478722291     i} & &\\
&\num{-7/2}&&\num{-0.47676177170605344 - 0.4998447478722291   i} & &\\
&&&&&\\

\multirow{4}*{149.}&\num{-5/2} &$\num{-5/2}_1$& \num{-0.5939692154358792} &\multirow{4}*{\num{-2.1321091791059117} }&
\multirow{4}*{\ytableausetup{centertableaux}
\begin{ytableau}
\none[4]&  & &\none[3] \\
\none[6] & &\none[1] \\
\none[6] & &\none[6]
\end{ytableau}}\\
&\num{7/2}&$\num{7/2}_1$ & \num{1.0416681842895437} && \\
\hline
\end{tabular}
\end{table}

\begin{table}
 \renewcommand\thetable{4.2}
\caption{(Continued)}
\begin{tabular}{l@{\hskip 0.2in}r@{\hskip 0.2in}r@{\hskip 0.2in}l@{\hskip 0.2in}l@{\hskip 0.2in}l@{\hskip 0.2in}}
\hline\hline
&$J$& $I_n$ & \hspace{0.4in}$\lambda$& \hspace{0.4in} $E$ & Rigged Configuration\\
\hline

\multirow{4}*{161.}&\num{1/2} &$\num{1/2}_1$& \num{-0.017936527015429728} &\multirow{4}*{\num{-3.903327748845358} }&
\multirow{4}*{\ytableausetup{centertableaux}
\begin{ytableau}
\none[4]&  & &\none[4] \\
\none[6] & &\none[4] \\
\none[6] & &\none[5]
\end{ytableau}}\\
&\num{5/2}&$\num{5/2}_1$ & \num{0.32293661081589986} && \\
&\num{7/2}&\multirow{2}*{$\num{2}_2$}&\num{0.9370084683865056 + 0.5320955305576359  i} & &\\
&\num{9/2}&&\num{0.9370084683865056 - 0.5320955305576359   i} & &\\
&&&&&\\

\multirow{4}*{162.}&\num{-1/2} &$\num{-1/2}_1$& \num{0.017936527015429728} &\multirow{4}*{\num{-3.903327748845358} }&
\multirow{4}*{\ytableausetup{centertableaux}
\begin{ytableau}
\none[4]&  & &\none[0] \\
\none[6] & &\none[1] \\
\none[6] & &\none[2]
\end{ytableau}}\\
&\num{-5/2}&$\num{-5/2}_1$ & \num{-0.32293661081589986} && \\
\hline
\end{tabular}
\end{table}

\begin{table}
 \renewcommand\thetable{4.2}
\caption{(Continued)}
\begin{tabular}{l@{\hskip 0.2in}r@{\hskip 0.2in}r@{\hskip 0.2in}l@{\hskip 0.2in}l@{\hskip 0.2in}l@{\hskip 0.2in}}
\hline\hline
&$J$& $I_n$ & $\lambda$& \hspace{0.1in} $E$ & Rigged Configuration\\
\hline
\multirow{4}*{173.}&\num{1/2} &$\num{1/2}_1$& \num{-0.022893719091156173} &\multirow{4}*{\num{-3.267165953541398} }&
\multirow{4}*{\ytableausetup{centertableaux}
\begin{ytableau}
\none[4]&  & &\none[4] \\
\none[6] & &\none[4] \\
\none[6] & &\none[6]
\end{ytableau}}\\
&\num{7/2}&$\num{7/2}_1$ & \num{0.7150852238364274} && \\
&\num{7/2}&\multirow{2}*{$\num{2}_2$}&\num{0.7535873719335956 + 0.5119580377157499  i} & &\\
&\num{9/2}&&\num{0.7535873719335956 - 0.5119580377157499  i} & &\\
&&&&&\\

\multirow{4}*{174.}&\num{-1/2} &$\num{-1/2}_1$& \num{0.022893719091156173} &\multirow{4}*{\num{-3.267165953541398} }&
\multirow{4}*{\ytableausetup{centertableaux}
\begin{ytableau}
\none[4]&  & &\none[0] \\
\none[6] & &\none[0] \\
\none[6] & &\none[2]
\end{ytableau}}\\
&\num{-7/2}&$\num{-7/2}_1$ & \num{-0.7150852238364274} && \\
\hline
\end{tabular}
\end{table}

\begin{table}
 \renewcommand\thetable{4.2}
\caption{(Continued)}
\begin{tabular}{l@{\hskip 0.2in}r@{\hskip 0.2in}r@{\hskip 0.2in}l@{\hskip 0.2in}l@{\hskip 0.2in}l@{\hskip 0.2in}}
\hline\hline
&$J$& $I_n$ & $\lambda$& \hspace{0.1in} $E$ & Rigged Configuration\\
\hline
\multirow{4}*{186.}&\num{1/2} &$\num{1/2}_1$& \num{0.17140368797514083} &\multirow{4}*{\num{-3.038187635447279} }&
\multirow{4}*{\ytableausetup{centertableaux}
\begin{ytableau}
\none[4]&  & &\none[0] \\
\none[6] & &\none[0] \\
\none[6] & &\none[3]
\end{ytableau}}\\
&\num{-7/2}&$\num{-7/2}_1$ & \num{-0.7141549160257158} && \\
&\num{-9/2}&\multirow{2}*{$\num{-2}_2$}&\num{-0.788289012751133 + 0.5148807356608802   i} & &\\
&\num{-7/2}&&\num{-0.788289012751133 - 0.5148807356608802 i} & &\\
&&&&&\\

\multirow{4}*{187.}&\num{5/2} &$\num{5/2}_1$& \num{0.3111830047955775} &\multirow{4}*{\num{-2.7809515687775552} }&
\multirow{4}*{\ytableausetup{centertableaux}
\begin{ytableau}
\none[4]&  & &\none[4] \\
\none[6] & &\none[6] \\
\none[6] & &\none[6]
\end{ytableau}}\\
&\num{7/2}&$\num{7/2}_1$ & \num{0.7358407067445326} && \\
\hline
\end{tabular}
\end{table}

\begin{table}
\renewcommand\thetable{4.2}
\caption{(Continued)}
\begin{tabular}{l@{\hskip 0.2in}r@{\hskip 0.2in}r@{\hskip 0.2in}l@{\hskip 0.2in}l@{\hskip 0.2in}l@{\hskip 0.2in}}
\hline\hline
&$J$& $I_n$ & $\lambda$& \hspace{0.1in} $E$ & Rigged Configuration\\
\hline

\multirow{4}*{199.}&\num{-7/2} &$\num{-7/2}_1$& \num{-1.26671672780226} &\multirow{4}*{\num{-2.462218658878338} }&
\multirow{4}*{\ytableausetup{centertableaux}
\begin{ytableau}
\none[4]&  & &\none[4] \\
\none[6] & &\none[0] \\
\none[6] & &\none[4]
\end{ytableau}}\\
&\num{3/2}&$\num{3/2}_1$ & \num{0.17324123549941658} && \\
&\num{7/2}&\multirow{2}*{$\num{2}_2$}&\num{1.1187744692931507 + 0.5489207268386115         i} & &\\
&\num{9/2}&&\num{1.1187744692931507 - 0.5489207268386115        i} & &\\
&&&&&\\

\multirow{4}*{200.}&\num{7/2} &$\num{7/2}_1$& \num{1.26671672780226} &\multirow{4}*{\num{-2.462218658878338} }&
\multirow{4}*{\ytableausetup{centertableaux}
\begin{ytableau}
\none[4]&  & &\none[0] \\
\none[6] & &\none[2] \\
\none[6] & &\none[6]
\end{ytableau}}\\
&\num{-3/2}&$\num{-3/2}_1$ & \num{-0.17324123549941658} && \\
\hline
\end{tabular}
\end{table}

\begin{table}
\renewcommand\thetable{4.3}
\caption{Bethe roots for  $N=12, M=4$, There are $45$ solution with one $1$-string and one $3$-string}
\begin{tabular}{l@{\hskip 0.2in}r@{\hskip 0.2in}r@{\hskip 0.2in}l@{\hskip 0.2in}l@{\hskip 0.2in}l@{\hskip 0.2in}}
\hline\hline
&$J$& $I_n$ & $\lambda$& \hspace{0.1in} $E$ & Rigged Configuration\\
\hline

\multirow{4}*{211.}&\num{-1/2} &$\num{0}_1$& \num{0.004745172056794713} &\multirow{4}*{\num{-2.6680318431345222} }&
\multirow{4}*{\ytableausetup{centertableaux}
\begin{ytableau}
\none[4]&  & & &\none[2] \\
\none[8] & &\none[4] 
\end{ytableau}}\\
&\num{11/2}&\multirow{3}*{$\num{0}_3$}&\num{1.0004857825077909       i} & &\\
&\num{1/2}&&\num{-0.004745172056794713} & &\\
&\num{-11/2}&&\num{-1.0004857825077909      i} & &\\
&&&&&\\

\multirow{4}*{212.}&\num{-1/2} &$\num{-1}_1$& \num{-0.14596813961790084} &\multirow{4}*{\num{-2.5094836538755265} }&
\multirow{4}*{\ytableausetup{centertableaux}
\begin{ytableau}
\none[4]&  & & &\none[2] \\
\none[8] & &\none[3] 
\end{ytableau}}\\
&\num{11/2}&\multirow{3}*{$\num{0}_3$}&\num{0.0330408101416682 + 1.0000750922972554        i} & &\\
\hline
\end{tabular}
\end{table}

\begin{table}
\renewcommand\thetable{4.3}
\caption{(Continued) }
\begin{tabular}{l@{\hskip 0.2in}r@{\hskip 0.2in}r@{\hskip 0.2in}l@{\hskip 0.2in}l@{\hskip 0.2in}l@{\hskip 0.2in}}
\hline\hline
&$J$& $I_n$ & $\lambda$& \hspace{0.1in} $E$ & Rigged Configuration\\
\hline

\multirow{4}*{224.}&\num{-1/2} &$\num{-1}_1$& \num{-0.17780173633522306} &\multirow{4}*{\num{-2.365118359195847} }&
\multirow{4}*{\ytableausetup{centertableaux}
\begin{ytableau}
\none[4]&  & & &\none[3] \\
\none[8] & &\none[3] 
\end{ytableau}}\\
&\num{9/2}&\multirow{3}*{$\num{1}_3$}&\num{0.5315009677745466 + 0.9996387074542886    i} & &\\
&\num{5/2}&&\num{0.5330073472673609} & &\\
&\num{11/2}&&\num{0.5315009677745466 - 0.9996387074542886     i} & &\\
&&&&&\\

\multirow{4}*{225.}&\num{1/2} &$\num{1}_1$& \num{0.17780173633522306} &\multirow{4}*{\num{-2.365118359195847} }&
\multirow{4}*{\ytableausetup{centertableaux}
\begin{ytableau}
\none[4]&  & & &\none[1] \\
\none[8] & &\none[5] 
\end{ytableau}}\\
&\num{-11/2}&\multirow{3}*{$\num{-1}_3$}&\num{-0.5315009677745466 + 0.9996387074542886    i} & &\\
\hline
\end{tabular}
\end{table}

\begin{table}
\renewcommand\thetable{4.3}
\caption{(Continued) }
\begin{tabular}{l@{\hskip 0.2in}r@{\hskip 0.2in}r@{\hskip 0.2in}l@{\hskip 0.2in}l@{\hskip 0.2in}l@{\hskip 0.2in}}
\hline\hline
&$J$& $I_n$ & $\lambda$& \hspace{0.1in} $E$ & Rigged Configuration\\
\hline

\multirow{4}*{236.}&\num{-7/2} &$\num{-4}_1$& \num{-1.299609203496181} &\multirow{4}*{\num{-0.8121317574768899} }&
\multirow{4}*{\ytableausetup{centertableaux}
\begin{ytableau}
\none[4]&  & & &\none[3] \\
\none[8] & &\none[0] 
\end{ytableau}}\\
&\num{9/2}&\multirow{3}*{$\num{1}_3$}&\num{0.6621506975137389 + 1.0013273523385884    i} & &\\
&\num{5/2}&&\num{0.6641750145094804} & &\\
&\num{11/2}&&\num{0.6621506975137389 - 1.0013273523385884         i} & &\\
&&&&&\\

\multirow{4}*{237.}&\num{7/2} &$\num{4}_1$& \num{-1.299609203496181} &\multirow{4}*{\num{-0.8121317574768899} }&
\multirow{4}*{\ytableausetup{centertableaux}
\begin{ytableau}
\none[4]&  & & &\none[1] \\
\none[8] & &\none[8] 
\end{ytableau}}\\
&\num{-11/2}&\multirow{3}*{$\num{-1}_3$}&\num{0.6621506975137389 + 1.0013273523385884    i} & &\\
\hline
\end{tabular}
\end{table}

\begin{table}
\renewcommand\thetable{4.3}
\caption{(Continued) }
\begin{tabular}{l@{\hskip 0.2in}r@{\hskip 0.2in}r@{\hskip 0.2in}l@{\hskip 0.2in}l@{\hskip 0.2in}l@{\hskip 0.2in}}
\hline\hline
&$J$& $I_n$ & $\lambda$& \hspace{0.1in} $E$ & Rigged Configuration\\
\hline

\multirow{4}*{249.}&\num{-7/2} &$\num{-3}_1$& \num{-0.4397065648183353} &\multirow{4}*{\num{-1.5633332262686777} }&
\multirow{4}*{\ytableausetup{centertableaux}
\begin{ytableau}
\none[4]&  & & &\none[0] \\
\none[8] & &\none[1] 
\end{ytableau}}\\
&\num{-11/2}&\multirow{3}*{$\num{-2}_3$}&\num{-0.9405176916828035 + 1.008858124558948          i} & &\\
&\num{-7/2}&&\num{-0.9949802258516587} & &\\
&\num{-9/2}&&\num{-0.9405176916828035 - 1.008858124558948          i} & &\\
&&&&&\\

\multirow{4}*{250.}&\num{-5/2} &$\num{-3}_1$& \num{-0.6811106701480948} &\multirow{4}*{\num{-1.0871519934870217} }&
\multirow{4}*{\ytableausetup{centertableaux}
\begin{ytableau}
\none[4]&  & & &\none[4] \\
\none[8] & &\none[1] 
\end{ytableau}}\\
&\num{9/2}&\multirow{3}*{$\num{2}_3$}&\num{1.1384664931814348 + 1.023578920970519           i} & &\\
\hline
\end{tabular}
\end{table}

\begin{table}
\renewcommand\thetable{4.4}
\caption{Bethe roots for $N=12, M=4$. There are $15$ solutions with two $2$-strings. $\epsilon^{12}=-6.551398760369255 \times 10^{-18 } $}
%
\end{table}

\begin{table}
\renewcommand\thetable{4.5}
\caption{Bethe roots  for $N=12$, $M=4$. There are $5$ solutions  with  one $4$-string}
\begin{tabular}{l@{\hskip 0.2in}r@{\hskip 0.2in}r@{\hskip 0.2in}l@{\hskip 0.2in}l@{\hskip 0.2in}l@{\hskip 0.2in}}
\hline\hline
&$J$& $I_n$ & $\lambda$& \hspace{0.1in} $E$ & Rigged Configuration\\
\hline

\multirow{4}*{271.}&\num{9/2} &\multirow{4}*{$\num{0}_4$}& \num{1.502976465754898i} &\multirow{4}*{ \num{-0.5022246220319766 }} &
\multirow{4}*{\ytableausetup{centertableaux}
\begin{ytableau}
\none[4]&  && & &\none[2] 
\end{ytableau}}\\
&\num{5/2}&&\num{ 0.5i} & &\\
&\num{7/2}&&\num{ -0.5i} & &\\
&\num{-9/2} & &\num{-1.502976465754898i} &&\\
&&&&&\\

\multirow{4}*{272.}&\num{9/2} &\multirow{4}*{$\num{1}_4$}& \num{0.5457966235030903 + 1.491557923422934 i} &\multirow{4}*{ \num{-0.46530094759009183 }} &
\multirow{4}*{\ytableausetup{centertableaux}
\begin{ytableau}
\none[4]&  && & &\none[3] 
\end{ytableau}}\\
&\num{9/2}&&\num{0.5430584220094842 + 0.49999812418171136 i} & &\\
\hline
\end{tabular}
\end{table}

\begin{table}
 \renewcommand\thetable{5.1}
\caption{Bethe roots  for $N=12$, $M=5$. There are  $21$ solutions with five   $1$-strings}
%
\end{table}

\begin{table}
 \renewcommand\thetable{5.2}
\caption{Bethe roots  for $N=12$, $M=5$. Three  are $105$ solutions with $1$-strings   and one $2$-string. 
$\epsilon^{12}=  -1.4956763315625075 \times 10^{-18}$ }
%
\end{table}

\begin{table}
 \renewcommand\thetable{5.2}
\caption{(Continued)}
%
\end{table}

\begin{table}
 \renewcommand\thetable{5.2}
\caption{(Continued)}
%
\end{table}

\begin{table}
 \renewcommand\thetable{5.2}
\caption{(Continued)}
%
\end{table}

\clearpage

\begin{table}
 \renewcommand\thetable{5.2}
\caption{(Continued)}
%
\end{table}


\begin{table}
 \renewcommand\thetable{5.3}
\caption{Bethe roots  for $N=12$, $M=5$. There are $84$ solutions with two  $1$-strings  and one $3$-string}
%
\end{table}

\begin{table}
 \renewcommand\thetable{5.3}
\caption{(Contunued)}
%
\end{table}

\begin{table}
 \renewcommand\thetable{5.3}
\caption{(Contunued)}
%
\end{table}

\clearpage


\begin{table}
 \renewcommand\thetable{5.3}
\caption{(Contunued)}
%
\end{table}

\begin{table}
 \renewcommand\thetable{5.3}
\caption{(Contunued)}
%
\end{table}

\begin{table}
 \renewcommand\thetable{5.3}
\caption{(Contunued)}
%
\end{table}

\begin{table}
 \renewcommand\thetable{5.4}
\caption{Bethe roots  for  $N=12$, $M=5$. There are $27$ solutions with  one $1$-string  and one $4$-string. $\epsilon^{12}
= 4.0574951652190555 \times 10^{-20}$}
%
\end{table}

\begin{table}
 \renewcommand\thetable{5.5}
\caption{Bethe roots  for $N=12, M= 5$. There are  $3$ solutions with one $5$-string}
%
\end{table}

\begin{table}
 \renewcommand\thetable{5.6}
\caption{Bethe roots  for $N=12, M= 5$. There are $42$ solutions with one $1$-string and two $2$-strings. $\epsilon^{12}= 7.400738616803484 \times 10^{-23}$}
%
\end{table}
\clearpage

\begin{table}
 \renewcommand\thetable{5.6}
\caption{(Continued)}
%
\end{table}

\begin{table}
 \renewcommand\thetable{5.7}
\caption{Bethe roots   for $N=12$, $M=5$. There is $15$  solutions with  one $2$-string  and one $3$-string}
%
\end{table}

\begin{table}
 \renewcommand\thetable{6.1}
\caption{Bethe root   for $N=12$, $M=6$. There is  $1$ solution with six    $1$-strings}
%
\end{table}

\begin{table}
 \renewcommand\thetable{6.2}
\caption{Bethe roots  for $N=12$, $M=6$. There are $15$ solutions with four $1$-strings and one $2$-string}
%
\end{table}

\begin{table}
 \renewcommand\thetable{6.3}
\caption{Bethe roots   for $N=12$, $M=6$. There are $35$ solutions with three $1$-strings and one $3$-string}
%
\end{table}

\begin{table}
 \renewcommand\thetable{6.4}
 \caption{Bethe roots  for $N=12, M=6$.  There are $28$ solutions with two $1$-strings   and one $4$-string. 
$\epsilon^{12}= 3.6999169395806464 \times 10^{-17}$}
%
\end{table}

\begin{table}
 \renewcommand\thetable{6.5}
\caption{Bethe roots  for $N=12, M= 6$.  There are $9$ solutions with  one $1$-string and one $5$-string}
%
\end{table}

\begin{table}
 \renewcommand\thetable{6.6}
\caption{Bethe root for $N=12, M= 6$. There is one solution with one $6$-string}
%
\end{table}

\begin{table}
 \renewcommand\thetable{6.7}
\caption{Bethe roots  for $N=12$, $M=6$. There are $15$ solutions with  two $1$-strings and  two  $2$-string}
%
\end{table}

\clearpage

\begin{table}
 \renewcommand\thetable{6.8}
\caption{Bethe roots  for $N=12$, $M=6$.  There are $21$ solutions with one  $1$-string, one  $2$-string  and one $3$-string}
%
\end{table}

\begin{table}
 \renewcommand\thetable{6.8}
\caption{(Continued)}
%
\end{table}

\clearpage

\begin{table}
 \renewcommand\thetable{6.8}
\caption{(Continued)}
%
\end{table}

\begin{table}
 \renewcommand\thetable{6.9}
\caption{Bethe root for $N=12, M=6$, There is one solution with   three $2$-strings}
%
\end{table}

\begin{table}
 \renewcommand\thetable{6.10}
\caption{Bethe roots  for $N=12, M=6$, There are $5$  solutions with  one $2$-string and one $4$-string}
%
\end{table}

\begin{table}
 \renewcommand\thetable{6.10}
\caption{(Continued)}
%
\end{table}

\section{Acknowledgement} 
The present study is partially supported by Grant-in-Aid for Scientific Research No. 24540396.
P. R. Giri acknowledges the financial support from JSPS.

\end{document}